\documentclass[a4paper,11pt]{article}
\pdfoutput=1 

\usepackage{jcappub} 

\usepackage[T1]{fontenc} 

\title{\boldmath Emission of gravitational waves by superconducting cosmic strings}

\author[a,b,1]{I. Yu. Rybak,\note{Corresponding author.}}
\author[a,b,c]{L. Sousa}

\affiliation[a]{Centro de Astrofísica da Universidade do Porto, Rua das Estrelas, 4150-762 Porto, Portugal}
\affiliation[b]{Instituto de Astrofísica e Ciências do Espaço, CAUP, Rua das Estrelas, 4150-762 Porto, Portugal}
\affiliation[c]{Departamento de F\'{\i}sica e Astronomia, Faculdade de Ci\^encias, Universidade do Porto, Rua do Campo Alegre 687, PT4169-007 Porto, Portugal}

\emailAdd{Ivan.Rybak@astro.up.pt}
\emailAdd{Lara.Sousa@astro.up.pt}

\abstract{We study the gravitational radiation emission efficiency $\Gamma$ of superconducting cosmic strings. We demonstrate, by using a solvable model of transonic strings, that the presence of a current leads to a suppression of the gravitational emission of cusps, kinks and different types of loops. We also show that, when a current is present, the spectrum of emission of loops with cusps is exponentially suppressed as the harmonic mode increases, thus being significantly different from the power law spectrum of currentless loops. Furthermore, we establish a phenomenological relationship between $\Gamma$ and the value of the current on cosmic strings. We conjecture that this relation should be valid for an arbitrary type of current-carrying string. We use this result to study the potential impact of current on the stochastic gravitational wave background generated by cosmic strings with additional degrees of freedom and show that both the amplitude and shape of the spectrum may be significantly affected. }

\begin{document}
\maketitle
\flushbottom

\section{Introduction}

Cosmic strings are one-dimensional topological defects, whose existence in cosmology was suggested by Tom Kibble~\cite{Kibble}. If these defects were created during a phase transition in the early universe, they would, in general, survive until the present time and thus the study of their observational signatures would shed light on several models of high-energy physics. As matter of fact, brane and hybrid inflation~\cite{SarangiTye, FirouzjahiTye, LazaridesPeddieVamvasakis, JonesStoicaTye, ChernoffTye}, supersymmetric grand unified theories~\cite{DavisDavisTrodden, JeannerotRocherSakellariadou, CuiMartinMorrisseyWells, Allys, Allys2}, axion models~\cite{DAVIS1986225, DABHOLKAR1990815, GorghettoHardyVilladoro, KawasakiKen'ichiSekiguchi}, the seesaw mechanism~\cite{PhysRevLett.124.041804, Samanta2021} and many other scenarios could be validated by
presence of cosmic strings.

The cosmic microwave background~\cite{LazanauShellard, LazanuShellardLandriau, LizarragaUrrestillaDaverioHindmarshKunz, CharnockAvgoustidisCopelandMoss, RybakAvgoustidisMartins, RybakSousa} (CMB) and lensing~\cite{Sazhin1, Sazhin2} have been used to probe the existence of cosmic strings and constrain the energy scale of the phase transition that may have originated them. More recently, however, considerable progress in constraining string-forming scenarios has been made by studying the gravitational wave signatures of cosmic strings~\cite{SousaAvelino, Blanco-PilladoOlum, RingevalSuyama, SousaAvelino2, LISA}. Although these studies have mostly focused on standard cosmic strings, more realistic scenarios, which include possible couplings with other fields, suggest that cosmic strings should be current-carrying~\cite{Babul:1987me, Everett, DavisPerkins, Peter:1993tm, GaraudVolkov, DavisPeter, Lilley:2010av, Fukuda:2020kym, AbeHamadaYoshioka}. In this paper we will bridge this gap, by studying the gravitational wave emission of superconducting cosmic strings.

Although for current-carrying strings, there may be an additional observational channel --- electromagnetic radiation~\cite{BlancoPilladoOlum, MiyamotoNakayama, Imtiaz2020} --- , the current traveling along the cosmic strings may be coupled with the hidden sector~\cite{BlinnikovKhlopov, SazhinKhlopov, BerezinskyVilenkin, HydeLongVachaspati, LongHydeVachaspati, LongVachaspati} and they may not generate electromagnetic signals as a result. The emission of gravitational radiation, on the other hand, should occur independently of the nature of the current and may then allow us to study a large variety of string-forming scenarios. Performing an accurate characterization of the observational signatures of superconducting cosmic strings, however, is challenging: there are no full superconducting cosmic strings simulations and the cosmological evolution of superconducting string networks is still unclear. However, recent progress in the semi-analytical description of the dynamics of networks of current-carrying strings~\cite{MPRS, MPRS2} allows for a better understanding of how their macroscopic evolution differs from that of standard cosmic string networks.  

In this paper, we study the gravitational radiation emission efficiency $\Gamma$ for current-carrying cosmic strings --- a quantity which is essential to accurately characterize their gravitational wave signatures. In particular, we investigate the gravitational wave emission of cusps, kinks and different types of loops, which allows us to establish a phenomenological relation between $\Gamma$ and the amplitude of the current that propagates along the cosmic strings. This result, coupled with the semi-analytical model to describe the evolution of current-carrying strings developed in~\cite{MPRS, MPRS2}, is then used to perform a preliminary study of the impact of current on the stochastic gravitational wave background. 

This paper is organized as follows. In section \ref{sec:transonic} we briefly review the dynamics of transonic superconducting strings. In section \ref{sec:gamma}, we introduce the framework that we will use to compute the gravitational radiation emission efficiency of superconducting cosmic strings. We then characterize the gravitation wave bursts emitted by cusp-like points on superconducting strings in section \ref{Cusp-like point} and study the impact of current on $\Gamma$ and the spectrum of emission of Burden loops in section \ref{sec:burden}. In section \ref{sec:kink}
we characterize the emission of kinks on strings with current and we compute the gravitational wave emission efficiency for cuspless superconducting loops in section \ref{sec:cuspless}. In section \ref{sec:SGWB} we study the potential impact of current on the shape and amplitude of the stochastic gravitational wave background generated by current-carrying strings. We then conclude in section \ref{sec:conclusions}. Throughout this paper, Greek indices run over spacetime coordinates (from $0$ to $3$) while Latin indices `$i$, $k$' run over spatial coordinates only. Latin indices `$a$ - $d$' run over worldsheet coordinates (from $0$ and $1$).

\section{Transonic superconducting strings}\label{sec:transonic}

Superconducting cosmic strings may be described effectively by an infinitely-thin string with degrees of freedom propagating on the string worldsheet. Infinitely thin strings sweep, in spacetime, a $1+1$-dimensional worldsheet $X^{\mu} (\tau,\sigma)$, parametrized by a timelike parameter $\sigma_0=\tau$ and a spacelike parameter $\sigma_1=\sigma$. The induced worldsheet metric is given by 
\begin{equation}
\gamma_{ab} \equiv X^{\mu}_{,a} X^{\nu}_{,b} \eta_{\nu \mu}\,,
\end{equation}
where $,a$ represents a derivative with respect to parameter $\sigma^a$ and $\eta_{\nu \mu}$ is the Minkowski metric. Their dynamics may then be described by an action of the form:
\begin{equation}
\label{EffectiveAction}
S_{\text{w}} = S_{\text{eff}} + S_{\text{int}} + S_{\text{em}}\,,
\end{equation}
where 
\begin{equation}
\label{Eff}
S_{\text{eff}} = \mu_0 \int \sqrt{-\gamma} \mathcal{F}(\kappa) d^2 \sigma\,, 
\end{equation}
and
\begin{equation}
\label{In}
S_{\text{em}} = \frac{1}{16 \pi} \int \sqrt{-\eta} F_{\mu \nu} F^{\mu \nu} d^4 x\,, \qquad S_{\text{int}} = q \int A_{a} \varepsilon^{ab} \phi_{,b} \, d^2 \sigma \,,
\end{equation}
describe, respectively, the effective action of an elastic string, of the electromagnetic fields, and the coupling of the current carriers to electromagnetism. Here, $\mu_0$ is a constant defined by the symmetry breaking scale, $\gamma=\det(\gamma^{ab})$ and $\eta=\det(\eta^{ab})$. The worldsheet Lagrangian, $\mathcal{F}(\kappa)$, has an internal degree of freedom
\begin{equation}
\kappa \equiv \phi_{,a} \phi_{,b} \gamma^{ab}\,,  
\end{equation}
where $\phi$ is a scalar field describing the charge carriers that is confined to the worldsheet. Moreover, $F_{\mu \nu} = A_{\nu, \mu} - A_{\mu,\nu} $, is the electromagnetic tensor, $A_{a} = A_{\mu} X^{\mu}_{,a}$ is the electromagnetic potential, and $\varepsilon^{ab}$ is a Levi-Civita symbol. Notice that the gauge invariance $A_{\mu} \rightarrow A_{\mu} + \partial_{\mu} \phi $ of the action (\ref{EffectiveAction}) is guaranteed by $\varepsilon^{ab}$ since $\phi_{,a} = \phi_{,\mu} X^{\mu}_{,a} $.
 
The equations of motion for superconducting strings can be obtained by varying the action in eq.~(\ref{EffectiveAction}) with respect to $\phi$:
\begin{equation}
   \label{EqOfMotBPVOA}
     \mu_0 \partial_{a} \left[ \mathcal{T}^{ab}  X^{\mu}_{,b} \right] = F^{\mu}_{\; \; \; \nu} X^{\nu}_{,a} J^a \sqrt{-\gamma} ,\qquad \mu_0 \partial_{a} \left[ \sqrt{-\gamma} \gamma^{ab} \mathcal{F}^{\prime}_{\kappa} \phi_{,b} \right] = \frac{1}{4} F_{ab} \varepsilon^{ab}\,,
\end{equation}
where we have defined the stress-energy tensor and electric $2$-current on the worldsheet as
\begin{equation}
\begin{gathered}
\label{TJ}
\mathcal{T}^{ab}  =  \sqrt{-\gamma} \left( \gamma^{ab} \mathcal{F} + \theta^{ab} \right), \quad
J^a = q \frac{\varepsilon^{a b} \phi_{,b}}{\sqrt{-\gamma}}\,,
\end{gathered}
\end{equation}
and $\theta^{ab}=-2\mathcal{F}'_\kappa\gamma^{ac}\gamma^{bd}\phi_{,c}\phi_{,d}$ is the stress-energy tensor of the current carriers.

The first effective model for current-carrying strings, proposed in \cite{Witten84}, suggested a worldsheet Lagrangian of the form:
\begin{equation}
\label{Witten}
\mathcal{F}_{\text{W}}(\kappa) =  1 - \frac{1}{2 \mu_0 } \kappa\,.
\end{equation} 
Later studies, however, demonstrated that this effective action does not reproduce the properties of the original $U(1)\times U(1)$ superconducting string model. To solve this problem, a number of alternative models were suggested, with particular interest being devoted to the action of transonic
cosmic strings \cite{Carter:1989dp, CARTER1989, Carter90, Carter95, CarterPeter, Carter2000}. Elastic strings are generally characterized by having distinct tension $T$ and mass per unit length $U$. As a result, these strings have two types of perturbations: wiggles, or transverse perturbations, propagating along the string at a speed of $c_E^2=T/U$, and longitudinal perturbations known as woggles with a speed of $c_L^2=-dT/dU$. When the speeds of propagation of wiggles and woggles coincide and are subluminal --- i.e. if $c_E=c_L<1$ --- these strings are said to be transonic\footnote{For standard Nambu-Goto strings (which have no internal degrees of freedom), the tension and mass per unit length coincide. As a result, for these strings $c_E=c_L=1$.}. It was shown in refs.~\cite{Carter90, Carter95} that the worldsheet Lagrangian is of the form:
\begin{equation}
\label{Action}
\mathcal{F}(\kappa) =  \sqrt{1 - \kappa}, \quad \kappa \in (-\infty,1].
\end{equation} 
Transonic strings provide an effective description of wiggles~\cite{Vilenkin90, Carter90, Carter95}, 5-dimensional projected Kaluza-Klein cosmic strings~\cite{NielsenOlesen, Nielsen} and some particular limits of superconducting strings --- see ref.~\cite{Carter2} for details\footnote{In appendix \ref{Appendix} one can see the additional possibility of deriving the transonic action as a ``linear'' superconducting string model.}. The attention devoted to this model is, to great extent, due to its integrability and consequent existence of exact solutions~\cite{Carter90, CarterSteer}. In this study, we will take advantage of this integrability to characterize their gravitational emission (and generalize some results for 
current-carrying strings with an arbitrary equation of state).

\section{Gravitational radiation emitted by superconducting cosmic strings}\label{sec:gamma}

We will consider general transonic string loops of the form 
\begin{equation}
\begin{gathered}
    \label{StrSolution}
    X^{\mu} = \frac{T_\ell}{2 \pi} \left[ X_+^{\mu} \left( \sigma_+ \right) + X_-^{\mu} \left(\sigma_- \right) \right],\\
     \phi = \frac{T_\ell}{2 \pi} \left[ F_+ \left( \sigma_+ \right) + F_- \left(  \sigma_- \right) \right]\,, 
\end{gathered}
\end{equation}
where we have decomposed $X^\mu$ and $\phi$ into left- and right-moving modes labelled, respectively, by the subscripts `$+$' and `$-$', $X_\pm^0 = \sigma_\pm$, $\sigma_\pm =  \frac{\pi}{T_\ell}  (\tau \pm \sigma) $ and $T_\ell$ is the period of oscillation that roughly corresponds to the length of the loop $\ell$ (see ref.~\cite{RybakAvgoustidisMartins2, Rybak} for details of this parameterization). The norm of the vector $\textbf{X}_{\pm}^{\prime}$ and the value of the current are related by
\begin{equation}
         \textbf{X}_+^{\prime \, 2}(\sigma_+) = 1 - F_+^{\prime \, 2}(\sigma_+), \quad \textbf{X}_-^{\prime \, 2}(\sigma_-) = 1 - F_-^{\prime \, 2}(\sigma_-)\,, 
\end{equation}
and, from now on, whenever the subscripts `$+$' or `$-$' are also present, primes represent the derivatives with respect to the corresponding unique coordinate $\sigma_{\pm}$. The worldsheet swept out by the loop throughout one period corresponds to the parameter interval $\tau \in [0, T_{\ell} ]$ and $\sigma \in [0, 2 T_\ell]$, or equivalently to $\sigma_\pm \in [0, 2\pi]$ (see ref.~\cite{AllenShellard1992} for details).

The stress-energy tensor of a transonic string may be written in the following form
\begin{equation}
\label{SEtens}
T^{\mu \nu} = \mu_0 \int \delta^{(4)} \left( x^{\lambda} - X^{\lambda}(\tau, \sigma) \right) \eta^{a b} X_{,a}^{\mu} X_{,b}^{\nu} \;  d \sigma d \tau\,,
\end{equation}
where $\eta^{a b}$ is a 2-dimensional Minkowski metric. The average of the Fourier transform of the stress-energy tensor over one period may conveniently be written as
\begin{equation}
\begin{gathered}
\label{FSEtens}
\tilde{T}^{\mu \nu} (\omega, \textbf{k}) = \frac{1}{T_{\ell}} \int T^{\mu \nu} \text{e}^{ - i k_{\nu} X^{\nu}} \;  d^4 x = \frac{\mu_0}{2 T_{\ell}} I_{+}^{(\mu} I_{-}^{\nu)}\,,
\end{gathered}
\end{equation}
where  $I_{+}^{(\mu} I_{-}^{\nu)} = \frac{1}{2} \left( I^{\mu}_+ I_{-}^{\nu}  + I_{-}^{\nu} I_{+}^{\mu} \right) $ and 
\begin{equation}
\label{IpIm}
I^{\mu}_{\pm} = \frac{T_\ell}{\pi}\int_0^{2\pi} X_{\pm}^{\prime \mu} \text{e}^{- {{ \frac{i T_\ell }{2 \pi} }} k_{\nu} X_{\pm}^{\nu}}  d \sigma_{\pm}\,. 
\end{equation}
Here, $k^{\mu} = \omega_j  n_{(1)}^{\mu} $, $\omega_j$ is the angular frequency emitted in $j$-th mode of emission and  $n^{\mu}_{(1)}=(1,\textbf{n}_{(1)})$ is the unit vector along the direction connecting the source to the observer. We shall work in the ``co-rotating'' orthogonal frame $\left(\textbf{n}_{(1)},\textbf{n}_{(2)},\textbf{n}_{(1)}\right)$, defined by
\begin{equation}
\begin{gathered}
\label{n23}
\textbf{n}_{(1)} = \left( \cos \theta, \; \sin \theta \cos \varphi, \; \sin \theta \sin \varphi \right) \,, \\
\textbf{n}_{(2)} = \left( -\sin \theta, \cos \varphi \cos \theta, \sin \varphi \cos \theta  \right) \,, \\
\textbf{n}_{(3)} = \left( 0, \sin \varphi, -\cos \varphi  \right)\,,
\end{gathered}
\end{equation}
and, for now on, the subscript `$j$' indicates that we are considering the contribution of the $j$-th harmonic mode of the corresponding variable.

The power emitted in the form of gravitational radiation $P$ can be expressed as~\cite{AllenShellard1992,Durrer}
\begin{equation}
\begin{gathered}
\label{GammaEff}
P =G \mu^2_0 \Gamma, \quad  \Gamma \equiv \sum_{j=0}^{\infty} \Gamma_j = \sum_{j=0}^{\infty} \int  \omega_j^2  \frac{  \left| I^{2 2}_{j} - I^{3 3}_{j} \right|^2  + \left| I^{2 3}_{j} + I^{3 2}_{j} \right|^2 }{2^3 \pi {T_\ell^2}  } d \Omega,
\end{gathered}
\end{equation}
where $\Omega(\theta, \varphi)$ is the solid angle, $| \dots |$ is the absolute value and $\Gamma$ is the efficiency of gravitational wave emission that we have split into the contribution of each harmonic mode $j$, $\Gamma_j$, characterized by angular frequency of $\omega_j = 2 \pi j/T_\ell$. Moreover, we have introduced:
\begin{equation}
\label{Iij}
I^{2 3}_{j} = I^{i}_+ I^k_- n_{(2) i} n_{(3) k}\,.    
\end{equation}

\subsection{Cusp-like point}
\label{Cusp-like point}

Cusps are expected to be a general feature of loops without current unless kinks prevent their formation \cite{GarfinkleVachaspati}. For standard strings, cusps form at points in which $\textbf{X}_{,1}=0$,
that thus move at the speed of light ($\textbf{X}_{,0}^2=1$). In terms of the decomposition into right- and left-moving modes, the condition of formation of a cusp may be written as:
\begin{equation}
\label{CuspCond}
\textbf{X}_{+ \, \text{c}}^{\prime}=\textbf{X}_{- \, \text{c}}^{\prime} \,,
\end{equation}
where the subscript `c' is used to label the values of the variables at the cusp.

For a current-carrying loop, the cusp condition (\ref{CuspCond}) may not be satisfied: in particular, if $F_+^{\prime \, 2}$ and $ F_-^{\prime \, 2}$ are distinct, cusps cannot form. 
Note, however, that for $F_+^{\prime \, 2} = F_-^{\prime \, 2}$ --- i.e. if current is symmetrical ---, the condition in eq.~(\ref{CuspCond}) is verified. Although the shape of the cusp is similar to that of cusps of standard string if current is symmetrical, the velocity of the cusp is necessarily subliminal and given by $\textbf{X}_{,0}^2 = 1 - F_{\pm}^{\prime \, 2}$~\cite{Rybak}.

To characterize the gravitational wave bursts emitted by cusps of superconducting strings, we shall use the formalism introduced in refs.~\cite{SpergelPiranGoodman,BlancoPilladoOlum} to study their electromagnetic radiation. Refs.~\cite{DamourVilenkin2, DamourVilenkin} and ref.~\cite{BabichevDokuchaev2} studied the gravitational wave bursts emitted by standard and chiral (null current) cosmic strings. Here, we shall go beyond and evaluate bursts of radiation emitted in the presence of time-like or space-like currents. For transonic strings, we will perform this computation analytically.

Let us consider a loop with currents such that the cusp condition (\ref{CuspCond}) is satisfied as closely as possible, i.e., such that $\textbf{X}_{+ \, \text{c}}^{\prime} \parallel \textbf{X}_{- \, \text{c}}^{\prime}$. Similarly to what was done in refs.~\cite{SpergelPiranGoodman,BlancoPilladoOlum,DamourVilenkin2,BabichevDokuchaev2}, we can then expand $\textbf{X}^{\prime}_\pm$ near the cusp-like point (which we assume to be located at $\sigma_\pm=0)$:
\begin{equation}
\label{CuspSeries}
\textbf{X}^{\prime}_\pm  (\sigma_{\pm}) = \textbf{X}^{\prime}_{\pm \, \text{c}} + \textbf{X}^{\prime \prime}_{\pm \, \text{c}} \sigma_{\pm} + \frac{1}{2} \textbf{X}^{\prime \prime \prime}_{\pm \, \text{c}} \sigma^2_{\pm} +\mathcal{O}(\sigma_\pm^3)\,.
\end{equation}
The integrals in eq.~(\ref{IpIm}) contracted with vectors (\ref{n23}) --- necessary to compute the quantities in eq.~(\ref{Iij}) --- thus reduce to
\begin{equation}
\begin{gathered}
\label{Integrals}
I^{i}_{\pm} n_{(2) i} \approx \frac{T_\ell}{\pi}\int_{-\infty}^{\infty} \left( X_{\pm 2}^{\prime}  + X_{\pm 2}^{\prime \prime} \sigma_{\pm} \right) \exp \left[ - i j X^{\mu}_{\pm} n_{(1) \mu}   \right] d \sigma_{\pm}, \\
I^{i}_{\pm} n_{(3) i} \approx \frac{T_\ell}{\pi} \int_{-\infty}^{\infty}  X_{\pm 3}^{\prime \prime} \sigma_{\pm} \exp \left[ - i j X^{\mu}_{\pm} n_{(1) \mu}   \right] d \sigma_{\pm},
\end{gathered}
\end{equation}
where $$X^{\mu}_{\pm} n_{(1)\mu} \approx (1 - X^{\prime}_{\pm 1}) \sigma_{\pm} - \frac{X^{\prime \prime}_{\pm 1}}{2}  \sigma_{\pm}^2 - \frac{X^{\prime \prime \prime}_{\pm 1}}{6} \sigma_{\pm}^3$$
and we have introduced $X^{(p)}_{\pm k} = \textbf{X}^{(p)}_{\pm \text{c}} \cdot \textbf{n}_{(k)}$. Since we anticipate that the vicinity of the cusp-like point emits most of the radiation, we have extended the limits of integration from $-\infty$ to $\infty$ (as was done in ref.~\cite{DamourVilenkin}).

The integrals in eq.~(\ref{Integrals}) can be expressed in a closed form using the Airy function $\text{Ai}(..)$ and its derivative $\text{Ai}^{\prime}(..)$\footnote{Previous studies~\cite{BlancoPilladoOlum, SpergelPiranGoodman, BabichevDokuchaev2} have resorted to modified Bessel functions to express these integrals. Note, however, that these solutions are not valid when the argument $\zeta_{\pm}$ has a negative value.}, as shown in appendix \ref{Appendix2}:
\begin{equation}
\begin{gathered}
\label{Integrals2}
I^{i}_{\pm} n_{(2) i} \approx   K_{\pm}  \left(   \frac{ X_{\pm 2}^{\prime} X^{\prime \prime \prime}_{\pm 1} - X^{\prime \prime}_{\pm 2} X^{\prime \prime}_{\pm 1}}{X^{\prime \prime \prime}_{\pm 1}}  \text{Ai}(\zeta_{\pm}) - i \frac{X_{\pm 2}^{\prime \prime} \text{sgn}\left[ X^{\prime \prime \prime}_{\pm 1} \right]}{(j|X^{\prime \prime \prime}_{\pm 1} | / 2)^{1/3}} \text{Ai}^{\prime}(\zeta_{\pm}) \right), \\
I^{i}_{\pm} n_{(3) i} \approx - K_{\pm} \left( X_{\pm 3}^{\prime \prime} \frac{X^{\prime \prime}_{\pm 1}}{X^{\prime \prime \prime}_{\pm 1}} \text{Ai}(\zeta_{\pm}) + i \frac{X_{\pm 3}^{\prime \prime} \text{sgn} \left[ X^{\prime \prime \prime}_{\pm 1} \right]}{(j |X^{\prime \prime \prime}_{\pm 1} | / 2)^{1/3}} \text{Ai}^{\prime}(\zeta_{\pm})  \right),
\end{gathered}
\end{equation}
where $$K_{\pm} = \frac{2 T_\ell \text{e}^{ij d_{\pm}} }{(j |X_{\pm 1}^{\prime \prime \prime}| / 2 )^{1/3}},\qquad \zeta_{\pm} = - \frac{j}{|X^{\prime \prime \prime}_{\pm 1}|} \frac{ \frac{1}{2} X^{\prime \prime \, 2}_{\pm 1}  + (1 - X^{\prime}_{\pm 1}) X^{\prime \prime \prime}_{\pm 1}}{(j |X^{\prime \prime \prime}_{\pm 1}| / 2)^{1/3}}, $$ $$d_{\pm} = - \frac{ X^{\prime \prime}_{\pm 1} }{ (X^{\prime \prime \prime}_{\pm 1})^2 } \left( \frac{ (X^{\prime \prime}_{\pm 1} )^2 }{3}  + (1-X^{\prime}_{\pm 1}) X^{\prime \prime \prime}_{\pm 1}  \right). $$
Note that, if we set $F_\pm^\prime=0$ in eq.~(\ref{Integrals2}) and assume that $n^{\mu}_{(1)}$ coincides with the direction of the burst (i.e., if $X^{\prime}_{\pm 2} =0 $ and $X^{\prime \prime}_{\pm 1} =0$),  we recover the result obtained in ref.~\cite{DamourVilenkin} for (currentless) Nambu-Goto strings.

\begin{figure} [h!]
\begin{center}
\includegraphics[width=6in]{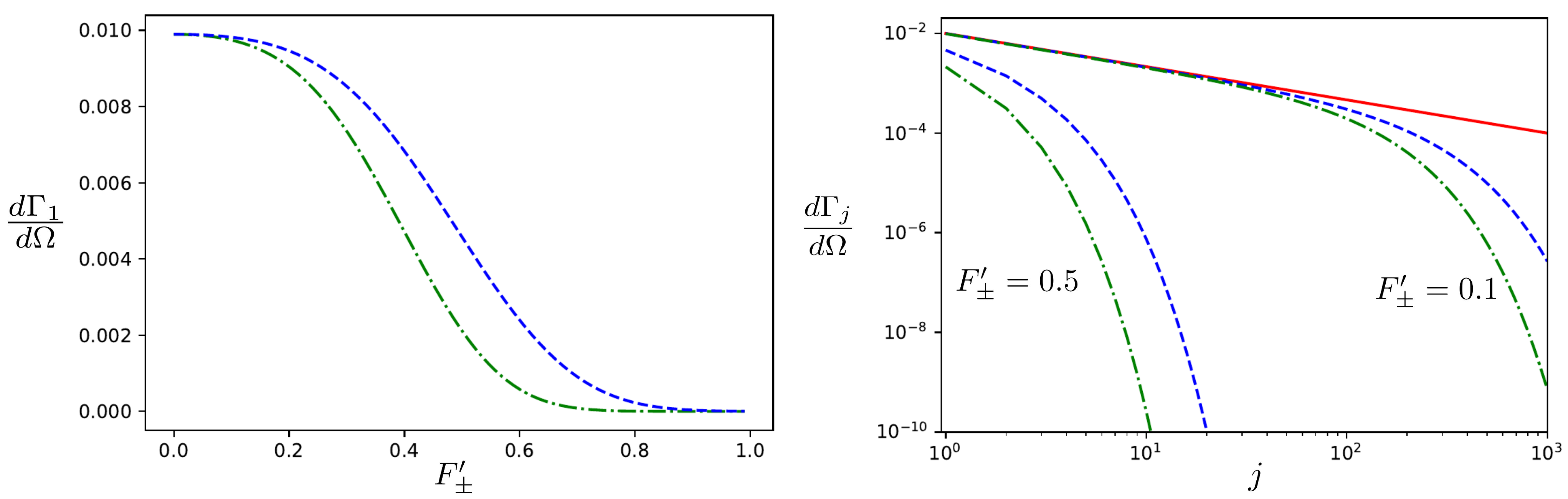}
\caption{\label{figure:CuspGamma} Gravitational radiation efficiency $\Gamma_j$ per unit solid angle $\Omega$ in the $j$-th harmonic mode of emission generated by a cusp. The left panel shows the dependence of $d\Gamma_1/d\Omega$ on the value of the current for symmetric (dash-dotted line) and chiral (dashed line) current-carrying strings. The right panel demonstrates the dependence of $d\Gamma_j/d\Omega$ on the harmonic mode of emission $j$ for symmetrical (dash-dotted line) and chiral (dashed line) current-carrying string for different values of the current as well as the power-law spectrum of Nambu-Goto strings (solid line).}
\end{center}
\end{figure}

To understand the impact that current has on the emission of gravitational radiation at a cusp-like point, let us evaluate the integrals in eq.~(\ref{Integrals2}) for a string configuration with constant $F^{\prime}_{\pm}$ (in other words, in a situation that is as close as possible to the standard cusp). This choice implies that
$\textbf{X}^{\prime}_{\pm} \parallel \textbf{n}_{(1)}$, 
$\textbf{X}^{\prime}_{\pm} \cdot \textbf{X}^{\prime \prime}_{\pm}=0$ and $\textbf{X}^{\prime \prime \prime}_{\pm} \cdot \textbf{X}^{\prime}_{\pm} = - \textbf{X}^{\prime \prime \, 2}_{\pm}$.
Hence, we should have that:
\begin{equation}
\begin{gathered}
\label{Integrals3}
I^{i}_{\pm} n_{(2) i} \approx  - i K_{\pm}  \frac{X_{\pm 2}^{\prime \prime} \text{sgn}\left[ X^{\prime \prime \prime}_{\pm 1} \right]}{(\omega |X^{\prime \prime \prime}_{\pm 1} | / 4)^{1/3}} \text{Ai}^{\prime}(\zeta_{\pm}), \\
I^{i}_{\pm} n_{(3) i} \approx - i K_{\pm} \frac{X_{\pm 3}^{\prime \prime} \text{sgn} \left[ X^{\prime \prime \prime}_{\pm 1} \right]}{(\omega |X^{\prime \prime \prime}_{\pm 1} | / 4)^{1/3}} \text{Ai}^{\prime}(\zeta_{\pm}) \,.
\end{gathered}
\end{equation}
By substituting these expressions into eq.~(\ref{GammaEff}), we may compute the gravitational wave emission efficiency (per unit solid angle) for superconducting strings. The results are plotted in figure~\ref{figure:CuspGamma} for different values of the current and different harmonic modes. Therein one can see that the efficiency of emission of gravitational radiation decreases significantly as the current increases. Moreover, as we increase the harmonic mode $j$, gravitational radiation emission efficiency decreases faster than the power-law observed for standard Nambu-Goto strings and this decay is faster for higher currents. Superconducting strings seem to have, then, a significantly different spectrum of emission. In the following subsection we shall compute the total gravitational wave emission efficiency $\Gamma$ for loops of superconducting strings with cusps and show that this is indeed the case.

\subsection{Burden's loops}\label{sec:burden}

In this section, we will study a particular class of loops, known as Burden's loops~\cite{BURDEN1985}. In this class, $\textbf{X}_\pm$ describe a circular motion in two planes oriented at an angle $\psi$:
\begin{equation}
\begin{gathered}
\label{ABmodesLoops1}
\textbf{X}_- = \frac{G_-}{N_-}\left(\cos{N_-\sigma_-},\sin{N_-\sigma_-},0\right) \,, \\
\textbf{X}_+ = \frac{G_+}{N_+}\left(\cos{N_+\sigma_+},\sin{N_+\sigma_+}\cos\psi,\sin{N_+\sigma_+}\sin\psi\right)\,,
\end{gathered}
\end{equation}
where $N_\pm$ are integers and relatively prime and $G_{\pm} = \sqrt{ 1 - F^{\prime \, 2}_{\pm}(\sigma_{\pm})}$ will be treated as constants. These loop solutions form cusps at discrete time instants and may thus be used to compute the gravitational wave power emitted by loops with cusps. They were, in fact, used to compute the gravitational radiation emitted by standard string loops~\cite{BURDEN1985} and chiral superconducting string loops~\cite{BabichevDokuchaev2002}. 

Here, we follow closely the procedure used in~\cite{BURDEN1985} in order to compute the power emitted by superconducting Burden loops using eq.~(\ref{GammaEff}). In this case, the integrals in eq.~(\ref{IpIm}), averaged over a string loop period, may be written as~\cite{BURDEN1985} 
\begin{equation}
\begin{gathered}
\label{IntegrBurden}
I^{\mu}_{\pm} = \frac{T_\ell}{\pi} \int_{0}^{2 \pi} \left[ 1, \; \textbf{X}^{\prime}_{ \pm }  \right]
\text{e}^{- i j \left( \sigma_{ \pm } - \textbf{n}_{(1)} \cdot \textbf{X}_{\pm} \right) }  d \sigma_{\pm}\,,
\end{gathered}
\end{equation} 
and computed using the relations in appendix \ref{Loops integrals}. Bearing in mind that $J_{n \pm 1}(z) = \frac{n}{z} J_n(z) \mp J_n^{\prime}(z)$, one can demonstrate that the final expression is similar to that of standard strings:
\begin{figure} [h!]
\begin{center}
\includegraphics[width=6in]{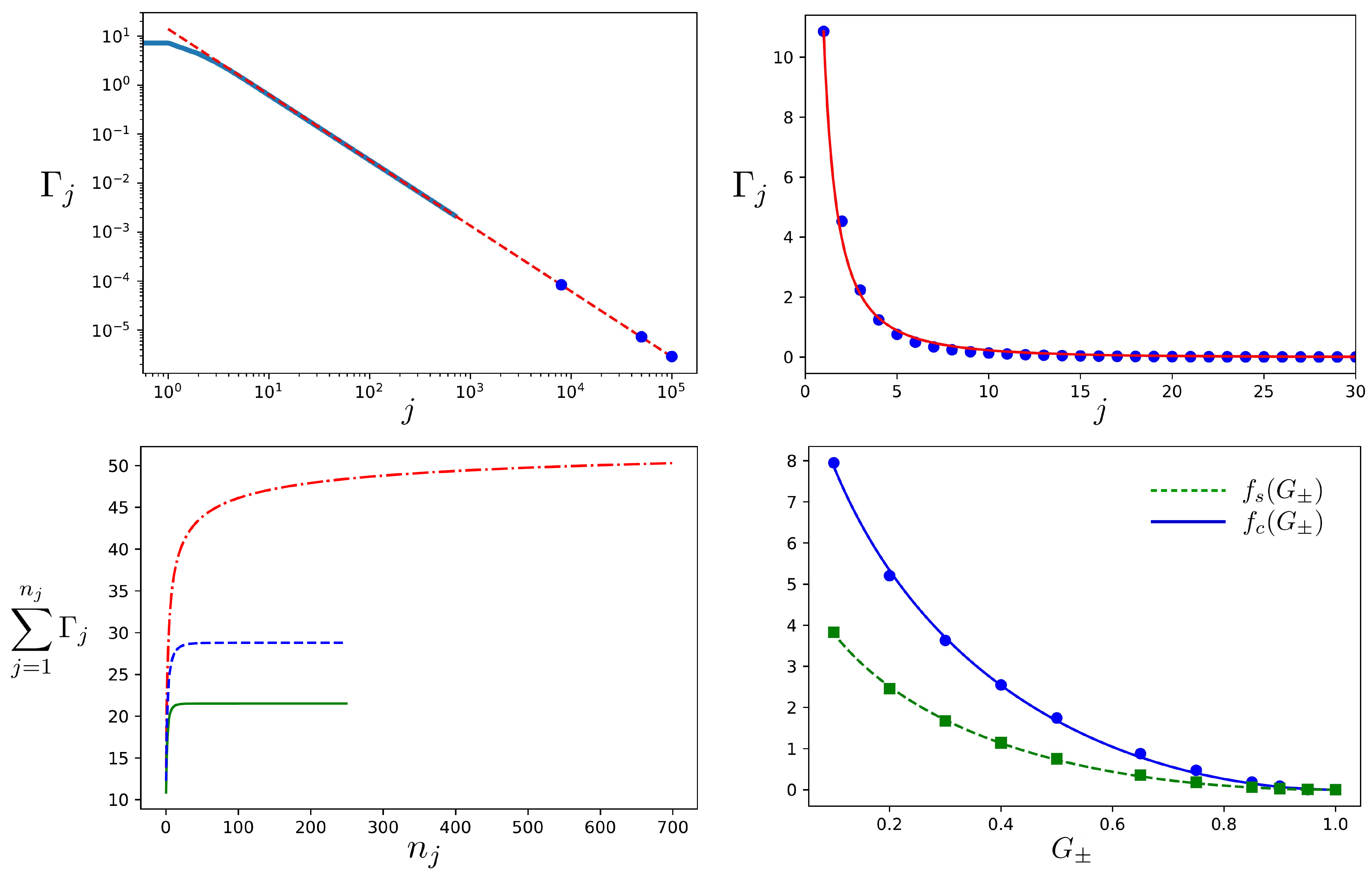}
\caption{\label{figure:BurdenTerms} 
Efficiency of gravitational radiation emission of loops with and without current. The left upper panel shows the gravitational radiation emission efficiency in each harmonic mode for standard cosmic strings. Therein, the solid line represents the first $700$ terms of $\Gamma_j$ and circles correspond to $j=800,50000,100000$, while the dashed line represents the power law approximation in eq.~(\ref{Power_q}). The fitted values of $q$ may be found in table~\ref{TableQ}. The right upper panel represents $\Gamma_j$ for different harmonic modes $j$ (circles) alongside the approximation in eq.~(\ref{Gammal_curr}) for superconducting loops with $G_{+}=G_{-}=0.9$. The bottom left panel represents the sum of first terms of $\Gamma_j$, $\sum_{j=1}^{N} \Gamma_j$, for Nambu-Goto loops with $G_+=G_-=1$ (dash-dotted line), loops with symmetrical current such that $G_+=G_-=0.9$ (solid line) and chiral loops with $G_+=1, \, G_-=0.9$ (dashed line) when $N_+=1$, $N_-=1$. The bottom right panel shows the behavior of the functions $f_{s,c}(G_{\pm})$ that appear in eq.~(\ref{Gammal_curr}), whose fitted parameters are listed in table~\ref{TableAB}. }
\end{center}
\end{figure}

\begin{equation}
\begin{gathered}
\label{GravRadAll}
\Gamma_j = 8 \pi j^2 \times \\ \int
 \left[ \left( s_+ s_- J_p J_q  + G_+ G_- J_p^{\prime} J_q^{\prime} \right)^2 + \left( s_+ G_- J_p J_q^{\prime}  + s_- G_+ J_q J_p^{\prime} \right)^2  \right] d \Omega,
\end{gathered}
\end{equation}
where $p=\frac{j}{N_+}$, $q=\frac{j}{N_-}$, $J_p = J_{p}(p \, G_+ d_+)$,
$J_q = J_{q}(q \, G_- d_-)$,
\begin{equation}
\begin{gathered}
\label{paramIntegr}
s_+ = \frac{\sin \theta \sin \varphi}{\sqrt{1-\sin^2 \theta \sin^2 \varphi}}\,, \quad
s_- = \frac{\sin \theta \sin (\psi-\varphi)}{\sqrt{1-\sin^2 \theta \sin^2 (\psi-\varphi)}}\,,\\
d_+ =  \cos\theta\left[1 + \tan^2 \theta \cos^2 \varphi\right]^{1/2}\,, \\
d_- =  \cos\theta\left[1 + \tan^2 \theta \cos^2(\varphi-\psi)\right]^{1/2} \,.
\end{gathered}
\end{equation}

\begin{figure} [h!]
\begin{center}
\includegraphics[width=3.8in]{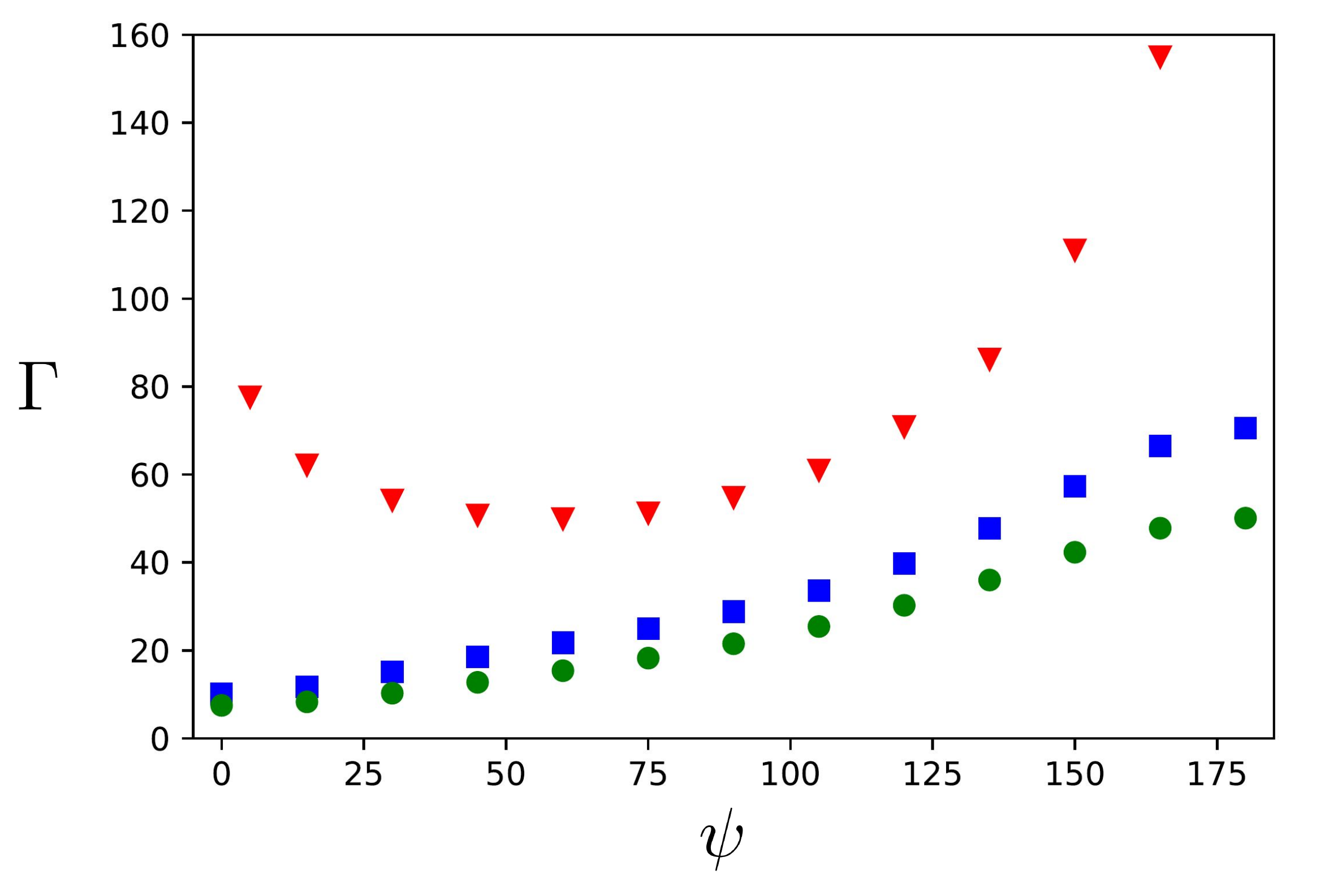}
\caption{\label{figure:BurdenDistrib} 
Values of the gravitational radiation emission efficiency $\Gamma$ for different values of $\psi$ for an ordinary loop with $G_+=G_-=1$ (triangles), a loop with symmetrical current $G_+=G_-=0.9$ (circles) and for a chiral loop $G_+=1, \, G_-=0.9$ (squares) when $N_{\pm}=1$. }
\end{center}
\end{figure}

\begin{figure} [h!]
\begin{center}
\includegraphics[width=6in]{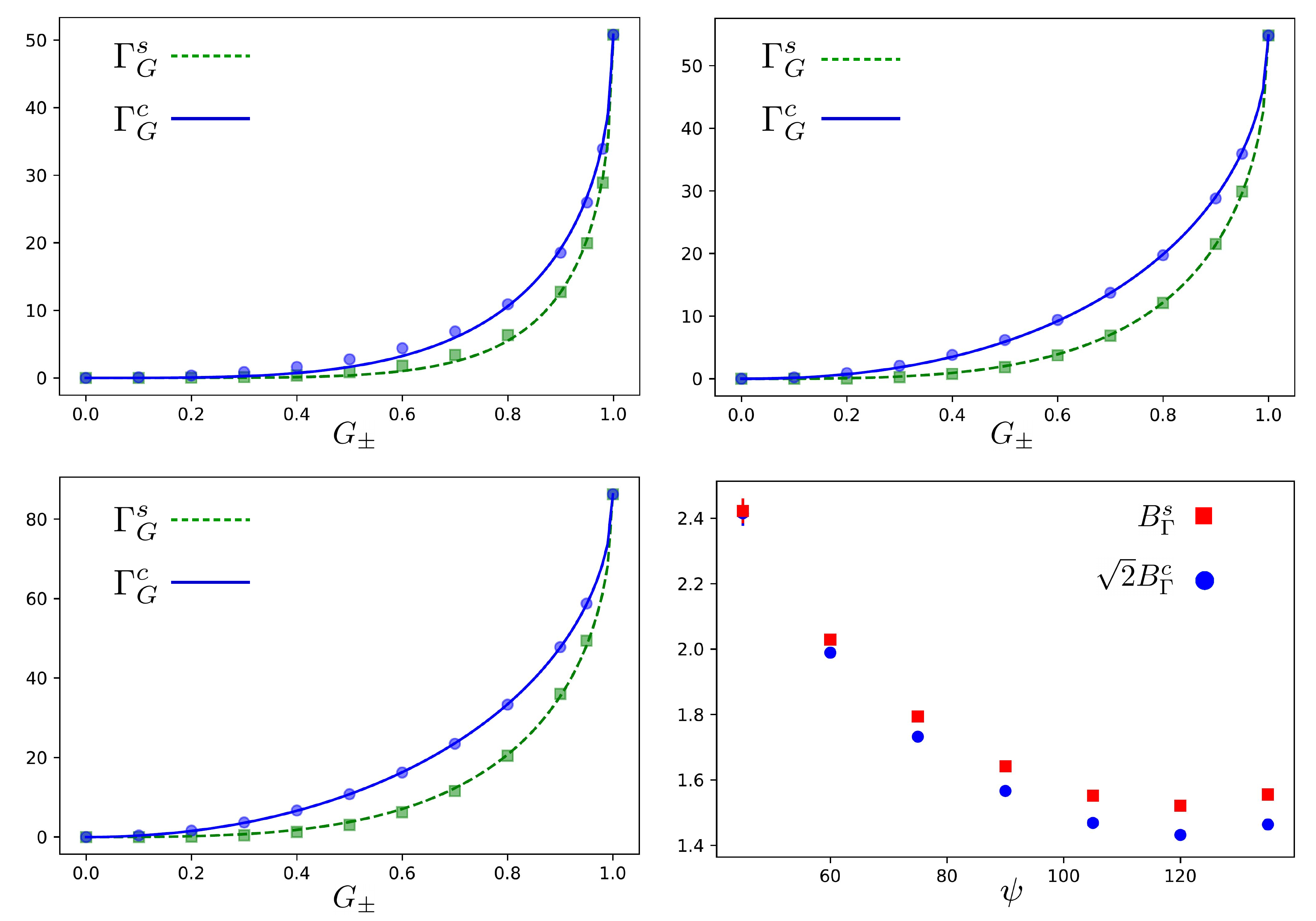}
\caption{\label{figure:Distrib}The gravitational radiation emission efficiency $\Gamma$ for different values of current and the fit by a function of the form of eq.~(\ref{Gamma_c_s}) for loops with symmetrical (squares and dashed lines) and chiral (circles and solid lines) currents  for different values of $\psi$: $45$ (upper left panel), 
$90$ (upper right panel) and $135$ (bottom left panel).
The bottom right panel represents values of the
fitted parameters $B^c_{\Gamma}$ (circles) and  $B^s_{\Gamma}$ (squares) with
corresponding errorbars, for different values of $\psi$. }
\end{center}
\end{figure}
Expression (\ref{GravRadAll}) represents explicitly the efficiency of emission of gravitational radiation $\Gamma$ of Burden loops, which can now be computed by numerical integration over angles $\theta$, $\varphi$ and by summing over the contributions of the different harmonic modes $j$. When $j$ increases, the contribution of each harmonic mode to the power radiated in the form of gravitational waves, $\Gamma_j$, decreases. For string loops without current, this decay assumes the form of a power law for large enough $j$:
\begin{equation}
\label{Power_q}
\Gamma_j \sim j^{-q}.
\end{equation} 
We can then perform the infinite summation over $j$ by summing over the first terms and by approximating the remaining terms using a power-law, as is shown in figure~\ref{figure:BurdenTerms}. The exponent of (\ref{Power_q}), or spectral index, is well described by $q \approx \frac{4}{3}$ (see table \ref{TableQ}), as it was argued in ref.~\cite{VachaspatiVilenkin}.

For cosmic string loops with current, each subsequent $\Gamma_j$ decays exponentially in the following form (by a similar argument as presented in ref.~\cite{VachaspatiVilenkin})
\begin{equation}
\label{Gammal_curr}
 \Gamma_j \sim j^{-q} \text{e}^{-j f_{m}(G_{\pm})}, \qquad 
  f_{m} (G_{\pm}) = a_{m} (1 - \sqrt{G_{\pm}})^{b_{m}},
\end{equation} 
where the labels $m=c$ and $m=s$ indicate that we considering strings with chiral or symmetrical currents, respectively. The fitted values of the parameters for both cases can be found in table~\ref{TableAB}. Since the power decreases faster with increasing $j$ when there is a current, in general one would need to consider a smaller number of terms in eq.~(\ref{GammaEff}) --- when compared to standard strings --- to accurately compute the total gravitational wave power $\Gamma$. This is clearly depicted in figure~\ref{figure:BurdenTerms}, which shows that, for $G_{\pm}=0.9$, considering only the first $250$ terms in eq.~(\ref{GammaEff}) already provides a good accuracy in the computation of $\Gamma$.
 
\begin{table} [h!]
\centering
\caption{Fitted spectral index $q$ of the power spectrum (eq.~(\ref{Power_q})) of Burden loops without current ($G_{\pm}=1$) with $N_{\pm}=1$.}
\label{TableQ}
  \begin{tabular}{ l | c | c | c | c | c | c | c | c | c | c  }
    \hline
   $\psi$  & $0 $  & $5 $ & $15$ & $30$ & $45, 60, ..., 150$ & $165$ & $175$ & $180$   \\ \hline
    $q$ & $0.99$ & $1.44$  & $1.40$ & $1.36$ &  $1.34$ & $1.40$ & $1.35$ & $1.00$  \\ \hline
  \end{tabular}
\caption{Fitted parameters of the exponential spectrum (eq.~(\ref{Gammal_curr})) for superconducting Burden loops with $N_{\pm} = 1$. }
\label{TableAB}
  \begin{tabular}{ l | c | c | c | c | c | c | c | c | c | c   }
    \hline
   $\psi$  & $0 $  & $15$ & $45$ & $60$ & $90$ & $120$ & $135$ & $165$  & $180$    \\ \hline
    $b_s$ & $2.09 $  & $2.28$ & $2.13$ & $2.04$ & $1.82$ & $1.67$ & $1.66$ & $1.73$ & $1.57$ \\ \hline
    $b_c$ & $2.28 $ & $2.70$ & $2.50$ & $2.35$ & $1.95$ & $1.65$ & $1.61$ & $1.71$ &  $1.44$ \\ \hline
        $\frac{1}{2} a_s$ & $8.20 $& $8.53 $  & $8.33$ & $8.19$ & $7.85$ & $7.55$ & $7.50$ & $7.60$ &   $7.32$ \\ \hline
    $a_c$ & $8.33 $  & $9.05$ & $8.76$ & $8.53$ & $7.97$ & $7.52$ & $7.46$ & $7.68$ &  $7.26$ \\ \hline
  \end{tabular}
\end{table}

Figure~\ref{figure:BurdenDistrib} shows $\Gamma$ for different values of $\psi$ for standard, chiral and current-symmetric loops. This figure clearly shows that current leads to a suppression of the gravitational wave power emitted by cosmic string loops. To better understand how $\Gamma$ changes in the presence of a current, we have studied the symmetric and chiral cases for different values of $\psi$ and performed a fit to the following expression
\begin{equation}
\label{Gamma_c_s}
\Gamma^{m}_{G} = \Gamma_{0} (1-|F^{\prime}_{\pm}|)^{B^{m}_\Gamma}, 
\end{equation} 
where $\Gamma_0$ is the gravitational radiation emission efficiency for loops without current. The results, including the fitted values of the parameters $B^{c}_{\Gamma}$ and $B^{s}_{\Gamma}$, for chiral and symmetrical currents, respectively, are shown in figure~\ref{figure:Distrib}. This figure clearly shows that $\Gamma$ decreases steeply as the current increases and that this suppression is more efficient for symmetrical currents when compared to chiral currents. As a matter of fact, one has that $B^{s}_{\Gamma} \approx \sqrt{2} B^{c}_{\Gamma}$.

\begin{table} [h!]
\centering
\centering
\caption{Fitted spectral index $q$ of the power spectrum (eq.~(\ref{Power_q})) of Burden loops without current ($G_{\pm}=1$) with $N_+ = 1$ and $N_-=3$.}
\label{TableQ3}
  \begin{tabular}{ l | c | c | c | c | c | c | c | c | c  }
    \hline
   $\psi$  & $15 $  & $30 $ & $45$ & $60$ & $75, 90, ..., 135$ & $150$ & $165$  \\ \hline
    $q$ & $1.46$ & $1.38$  & $1.36$ & $1.35$ &  $1.34$ & $1.35$ & $1.40$  \\ \hline
  \end{tabular}
\caption{Fitted parameters of the exponential spectrum (eq.~(\ref{Gammal_curr})) for superconducting Burden loops with $N_+=1$ and $N_-=3$. }
\label{TableAB3}
  \begin{tabular}{ c | c | c | c | c | c | c    }
    \hline
   $\psi$  & $b_{3s}$ & $b_{3c}$ & $ \frac{1}{\sqrt{2}} a_{3s}$ & $a_{3c}$ & $ B^s_{3\Gamma}$ & $2^{1/4} B^c_{3\Gamma}$ \\ \hline
   $45$ &  $1.99$ & $1.99$ & $23.37$ & $24.34 $ & $2.79 $ & $2.80 $ \\ \hline
   $90$ &  $1.74$ & $1.76$ & $21.68$ & $23.12$  & $2.06$ & $2.01$ \\ \hline
   $135$  & $1.76$ & $1.77$ & $21.63$ & $22.97$ & $2.14$ & $2.11$ \\ \hline
  \end{tabular}
\end{table}

\begin{figure} [h!]
\begin{center}
\includegraphics[width=6in]{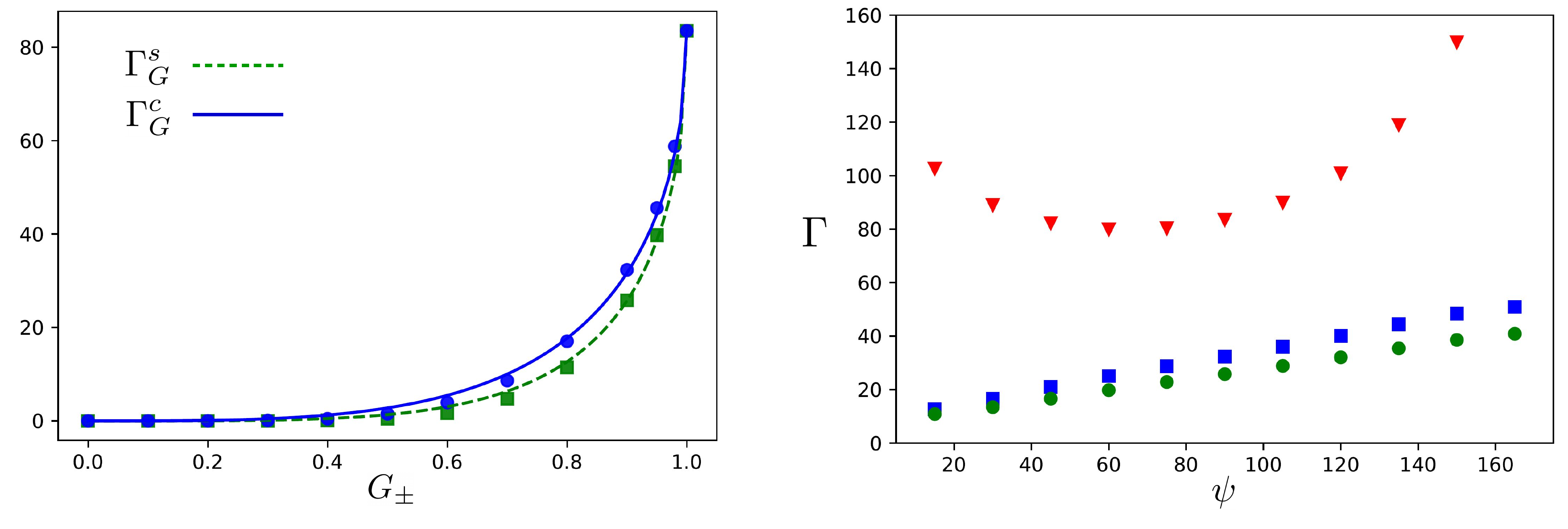}
\caption{\label{figure:Distrib3} Impact of current on the efficiency of emission of gravitational waves by Burden loops with $N_{+}=1$, $N_- = 3$. The left panel represents values of $\Gamma$ for different $\psi$ for an ordinary loop with $G_+=G_-=1$ (triangles), a loop with symmetrical current with $G_+=G_-=0.9$ (circles) and a chiral loop $G_+=1, \, G_-=0.9$ (squares) when $N_{+}=1$, $N_- = 3$. The right panel shows the best fit by eq.~(\ref{Gamma_c_s}) to the computed values of $\Gamma$, chiral loops (solid line and circles) and for loops with symmetrical current (dashed line and squares).}
\end{center}
\end{figure}

We have also carried out a similar analysis for Burden loops with $N_+=1$ and $N_-=3$. The impact of current on the power emitted in gravitational radiation is qualitatively very similar (as figure~\ref{figure:Distrib3} shows): current leads to an exponential power spectrum of the form of eq.~(\ref{Gammal_curr}) and to decrease of $\Gamma$ that is well described by an approximation of the form of~(\ref{Gamma_c_s}). The values of the fitted parameters --- recorded in tables~\ref{TableQ3}, \ref{TableAB3} --- are also quantitatively very similar and in good agreement with those found for loops with $N_\pm=1$.

\subsection{Radiation from kinks}\label{sec:kink}

Cosmic string collisions and self-intersections result in intercommutations and reconnections that produce discontinuities in the string tangent vector known as kinks. Kinks travel along the string at the speed of light for ordinary Nambu-Goto strings, but for superconducting strings their velocities are smaller --- see ref.~\cite{Rybak} for more details. As we shall see in this section, this necessarily has an impact on the gravitational wave burst emitted by the kink.

Note that one can always choose a region sufficiently close to the kink where the string segments are straight. In this case, for any function $\mathcal{F}(\kappa)$ in action (\ref{Eff}), the string solution is still of the form (\ref{StrSolution}), and the dynamics is described by an action of the form:
\begin{equation}
\begin{gathered}
\label{ActionJunctCurrent}
   S = - \mu_{0} \sum_m  \int \mathcal{F}(\kappa_m) \sqrt{-\gamma_m} \, \Theta \left( s_m(\tau) - \sigma_m \right)  d \sigma_m d \tau +\\
 \sum_m \int \mathrm{h}_{\mu m} \left( x_m^{\mu} (s_m(\tau),\tau) - \mathcal{X}^{\mu}(\tau) \right)  d \tau  
 +  \sum_m \int \mathrm{g}_m \left( \varphi_m (s_m(\tau),\tau) - \Phi(\tau) \right) d \tau, 
\end{gathered}
\end{equation}
which describes two connected string segments, labeled by $m=I$ and $m=II$, subjected to continuity conditions for the segments and their currents at the kink (enforced, respectively, by the last two terms). Here, $\mathrm{h}_{\mu m}$, $\mathrm{g}_m$ are Lagrange multipliers for strings and currents, functions $\mathcal{X}^{\mu}(\tau)$ and $\Phi(\tau)$ define the values of $x_m^\mu$ and $\phi_m$ at the point where the segments meet and function $s(\tau)$ describes the worldsheet coordinates $\sigma$ of the kink. 

The stress-energy tensor for the action (\ref{ActionJunctCurrent}) may be written as 
\begin{equation}
\begin{gathered}
\label{SEKink}
T^{\mu \nu}  =\frac{2}{\sqrt{-g}} \sum_m \int \mathrm{h}^{(\mu}_m \left( X_m^{\nu)} - \mathcal{X}^{\nu)} \right) \delta^{(4)}\left( x_n^{\lambda} - X^{\lambda} \right) d \tau +\\
+ \frac{1}{\sqrt{-g}} \mu_0 \sum_m  \int  \Theta(s_m - \sigma_m) \sqrt{-\gamma}_m \left( U_m u_m^{\mu} u_m^{\nu} - T_m v_m^{\mu} v_m^{\nu} \right)  \delta^{(4)}\left( x^{\lambda}_m - X^{\lambda} \right) d \sigma d \tau\,.
\end{gathered}
\end{equation}
Note that the first term vanishes as a result of the continuity condition --- $x^{\mu}_n (s_n, \tau) = \mathcal{X}^{\mu}(\tau)$ --- that emerges by varying the action (\ref{ActionJunctCurrent}) with respect  to $\mathrm{h}_{\mu m}$. The Fourier-transformed stress-energy tensor then has the form
\begin{equation}
\begin{gathered}
\label{SEKink22}
\tilde{T}^{\mu \nu} =  \frac{\mu_0 T_\ell}{2\pi^2}\int_{-\infty}^{\infty}  \Bigg[ \text{e}^{-i j X^{\lambda}_{I+} n_{(1) \lambda} } X_{I+}^{\prime  (\mu} \int_{-\infty}^{\sigma_{-}^s(\sigma_+)}  X_{I -}^{\prime \nu)} \text{e}^{-i j X_{I -}^{\lambda} n_{(1) \lambda} } d \sigma_- + \\
+ \text{e}^{-i j X_{II +}^{\lambda} n_{(1) \lambda} } X_{II+}^{\prime  (\mu} \int_{\sigma_{-}^s(\sigma_+)}^{\infty} X_{II -}^{\prime \nu)} \text{e}^{-i j X_{II-}^{\lambda} n_{(1) \lambda} } d \sigma_- \Bigg] \frac{d \sigma_+}{2},
\end{gathered}
\end{equation}
where we have changed the variables of integration from $\tau,\, \sigma$ into $\sigma_+$ and $\sigma_-$ (as defined after
eq.~(\ref{StrSolution})) and the function $\sigma_{-}^s (\sigma_+)$ is defined by
\begin{equation}
\label{Eq_for_s}
s \left( \frac{\sigma^s_+ + \sigma^s_- }{2} \right) - \frac{ \sigma^s_+ -  \sigma^s_- }{2}=0.
\end{equation}
Note that kinks correspond to a discontinuity in either the left- or the right-moving modes. Here, we have assumed, without loss of generality, that $X_-^\mu(\sigma_-)$ has a discontinuity at $\sigma_-^s$. This corresponds to kinks moving in the positive $s$ direction ($\dot{s}=1$), for which eq.~(\ref{Eq_for_s}) assumes the form: 
$$\sigma_k + \tau^s - \sigma^s =  0 \rightarrow \sigma_{-}^s = - \sigma_k,$$
where $\sigma_k$ is a constant defining location of a kink on the worldsheet. If the kink moves in the opposite direction ($\dot{s}=-1$), we should exchange the order of integration in eq.~(\ref{SEKink22}). Then, by performing the integration over $\sigma_-$ and keeping only the terms up to first order in the expansion, we find: 
\begin{equation}
\begin{gathered}
\label{SEKink4}
\tilde{T}^{\mu \nu} \approx i\frac{\mu_0 }{j} \frac{T_\ell}{(2\pi)^2} \Big[ 
 \frac{ X_{I-}^{\prime (\nu}}{1-X_{I-1}^{\prime} }
 -  \frac{ X_{II-}^{\prime (\nu}}{1- X_{II-1}^{\prime}  } \Big] 
 \int_{-\infty}^{\infty} X_{+}^{\prime  \mu)} \text{e}^{-i j (1 - X^{\prime}_{+1})\sigma_+}  d \sigma_+,
\end{gathered}
\end{equation}
where we have used the fact that the discontinuity appears only in the first derivative of $X^\mu_-$ to set $X^\mu_{I\pm}=X^\mu_{II\pm}=X^\mu_{\pm}$ and $X'^\mu_{I+}=X'^\mu_{II+}=X'^\mu_{+}$ and the integral over $\sigma_+$ should be carried as described in section \ref{Cusp-like point}. This expression is very similar to that found for Nambu-Goto strings~\cite{Vilenkin:2000jqa} and hence eq.~(\ref{SEKink4}) describes the gravitational radiation emitted by kinks with and without the current. The presence of a current, however, does not allow for the situation in which $X^{\mu}_{\pm} n_{\mu} \rightarrow 0$ and also leads to an increase of the value of the denominator. This always leads to a decrease in the gravitational radiation emission efficiency of current-carrying strings when compared to Nambu-Goto strings and, thus, one shall expect the gravitational wave bursts emitted by kinks of strings with current to be less pronounced (as is the case for cusps). In the next section, we shall consider a particular class of loops with kinks to demonstrate that this is indeed the case.

\subsection{Cuspless loops}\label{sec:cuspless}

Let us consider a loop consisting of four straight string segments connected by four kinks \cite{GarfinkleVachaspati2, CopelandHindmarshTurok}
\begin{equation}
\label{CuspLessLoop}
\textbf{X}_{\pm} = G_{\pm} \textbf{r}_{\pm} \begin{cases} \sigma_{\pm} - \frac{\pi}{2}, \qquad 0 \leq \sigma_{\pm} \leq \pi, \\ \frac{3 \pi}{2} -  \sigma_{\pm}, \qquad \pi \leq \sigma_{\pm} \leq 2 \pi,  \end{cases}
\end{equation}
where $\textbf{r}_{\pm}$ are unit vectors. Solution (\ref{CuspLessLoop}) is valid for any current-carrying string provided that $G_{\pm}$ are constants (which can be seen directly from the equations of motion in eq.~(\ref{EqOfMotBPVOA})). As a matter of fact, ref.~\cite{BabichevDokuchaev2002} investigated such loops for a chiral current.

\begin{figure} [h!]
\begin{center}
\includegraphics[width=6.in]{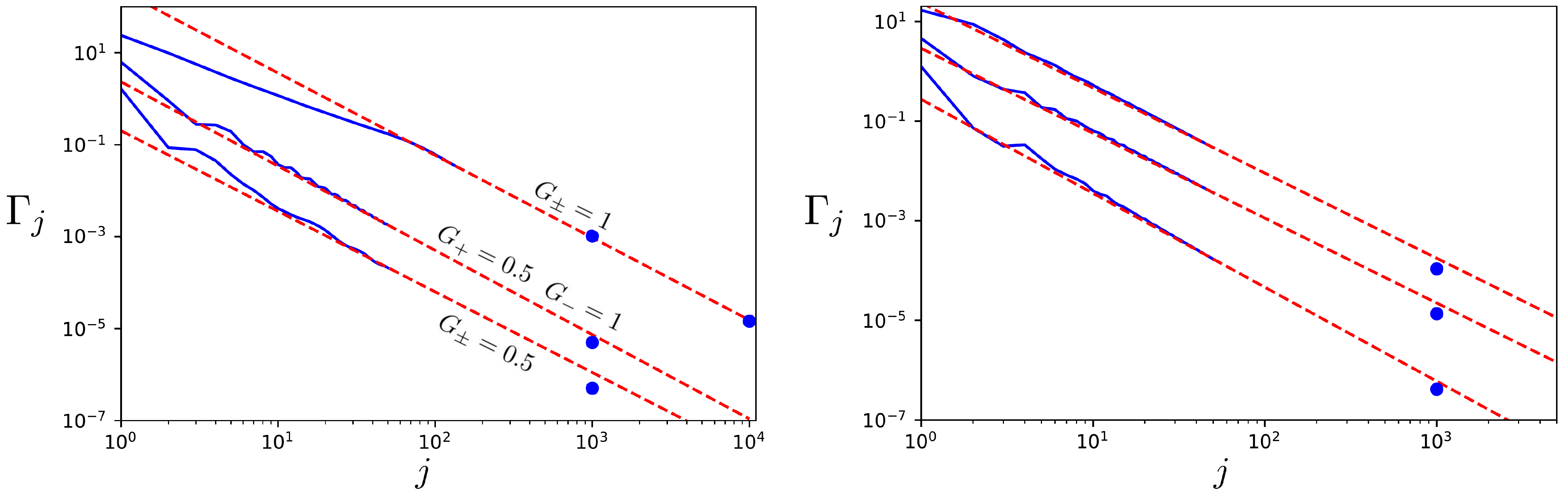}
\caption{\label{figure:CuspLessTerms} 
Power spectrum of cuspless loops. The left panel shows power-law approximation (dashed lines), the first terms of $\Gamma_j$ (solid lines) and values of $\Gamma_j$ for $j=1000$  (dots) for different values of $G_{\pm}$ and $\alpha = \frac{\pi}{12}$. The right panel represents the same quantities when $\alpha = \frac{\pi}{2}$.}
\end{center}
\end{figure}

To characterize the change in the efficiency of emission of gravitational radiation when current is not trivial, we follow a procedure similar to that used in section~\ref{sec:burden} and plug the solution in eq.~(\ref{CuspLessLoop}) into the integrals (\ref{FSEtens}). This yields
\begin{equation}
\label{IpmCuspLess}
\textbf{I}_{\pm}  = i \frac{2 T_\ell}{\pi} G_{\pm} \textbf{r}_{\pm} \frac{(-1)^j \text{e}^{i \frac{\pi}{2} j G_{\pm} \textbf{n}_{(1)} \cdot \textbf{r}_{\pm} } - \text{e}^{- i \frac{\pi}{2} j G_{\pm} \textbf{n}_{(1)} \cdot \textbf{r}_{\pm} }  }{j \left[ 1 - (G_{\pm} \textbf{n}_{(1)} \cdot \textbf{r}_{\pm} )^2 \right] } \equiv \textbf{r}_{\pm} \mathcal{C}_{\pm},
\end{equation}
where we will be using only spatial part.
One can see that 
\begin{equation}
\label{CCStar}
\mathcal{C}_{\pm} \mathcal{C}^*_{\pm} = 2 \left( \frac{2 T_\ell}{\pi} G_{\pm} \right)^2  \frac{1 - (-1)^j \cos \left( \pi j G_{\pm} \textbf{n}_{(1)} \cdot \textbf{r}_{\pm}  \right)}{j^2 \left[ 1 - \left( G_{\pm} \textbf{n} \cdot \textbf{r}_{\pm} \right)^2 \right)^2}\,,
\end{equation}
and thus
\begin{equation}
\label{Iqp}
I^{q p}_{+-} = \left( \textbf{n}_{(q)} \cdot \textbf{I}_+ \right) \left( \textbf{n}_{(p)} \cdot \textbf{I}_-\right) = \mathcal{C}_{+} \mathcal{C}_{-} \left( \textbf{n}_{(q)} \cdot \textbf{r}_+ \right) \left( \textbf{n}_{(p)} \cdot \textbf{r}_-\right)\,.
\end{equation}
Following the steps taken in refs.~\cite{GarfinkleVachaspati, BabichevDokuchaev2002}, we obtain an expression for $\Gamma_j$.
\begin{equation}
\begin{gathered}
\label{GammaCusplessL}
\Gamma_j = \frac{32}{\pi^3} \int  \frac{  G_{+}^2 G_-^2 \left( 1 - (\textbf{n}_{(1)} \cdot \textbf{r}_-)^2 \right) \left( 1 - (\textbf{n}_{(1)} \cdot \textbf{r}_+)^2 \right) }{\left( 1 - \left( G_{+} \textbf{n}_{(1)} \cdot \textbf{r}_{+} \right)^2 \right)^2 \left( 1 - \left( G_{-} \textbf{n}_{(1)} \cdot \textbf{r}_{-} \right)^2 \right)^2} \times \\
  \frac{\left(1 - (-1)^j \cos(\pi j G_+ \textbf{n}_{(1)} \cdot \textbf{r}_+)  \right) \left(1 - (-1)^j \cos(\pi j G_- \textbf{n}_{(1)} \cdot \textbf{r}_-)  \right) }{j^2}  d \Omega
 \end{gathered}
\end{equation}
and $\Gamma$
\begin{equation}
\begin{gathered}
\label{GammaCuspless}
\Gamma = \frac{16}{\pi} \int  \frac{  G_{+}^2 G_-^2 \left( 1 - (\textbf{n}_{(1)} \cdot \textbf{r}_-)^2 \right) \left( 1 - (\textbf{n}_{(1)} \cdot \textbf{r}_+)^2 \right) }{\left( 1 - \left( G_{+} \textbf{n}_{(1)} \cdot \textbf{r}_{+} \right)^2 \right)^2 \left( 1 - \left( G_{-} \textbf{n}_{(1)} \cdot \textbf{r}_{-} \right)^2 \right)^2} \times \\
 \left( 1 - \frac{1}{2} \left( | G_+ \textbf{n}_{(1)} \cdot \textbf{r}_+ + G_- \textbf{n}_{(1)} \cdot \textbf{r}_- | +  | G_+ \textbf{n}_{(1)} \cdot \textbf{r}_+ - G_- \textbf{n}_{(1)} \cdot \textbf{r}_- | \right) \right) d \Omega.
 \end{gathered}
\end{equation}
We chose a coordinate system such that
$$\textbf{n}_{(1)} \cdot \textbf{r}_{\pm} = \sin \theta \cos \left[ \alpha/2 \pm \varphi \right].$$
Carrying out the calculations numerically, one may see that each subsequent term of $\Gamma_j$ decays according to power-law (\ref{Power_q}) with $q\approx \frac{5}{3}$, for both standard and current-carrying cuspless loops. Figure~\ref{figure:CuspLessTerms} shows examples of the fitting and table~\ref{TableQKink} provides the corresponding fitted values.
\begin{table} [h!]
\centering
\centering
\caption{Fitted parameters of the power spectrum in eq.~(\ref{Power_q}) for cuspless loops and of the fitting function for $\Gamma$ in eq.~(\ref{Gamma_c_s}). Here $q_c$ and $B^c_{\Gamma}$ refer to the values of these parameters for chiral loops, while $q_s$ and $B^s_{\Gamma}$ to those of loops with $G_{+}=G_{-}$.}
\label{TableQKink}
  \begin{tabular}{ r | c | c | c | c | c | c | c   }
    \hline
   $\alpha$  & $15 $  & $45 $ & $90$ & $135$ & $165$   \\ \hline
    $(G_{\pm}=1) \; q $ & $1.79$ & $1.65$  & $1.71$ & $1.65$ &  $1.72$  \\ \hline
    $(G_{+}=0.9) \; q_c$ & $1.74$ & $1.69$  & $1.74$ & $1.69$ &  $1.74$  \\ \hline
    $(G_{+}=0.5) \; q_c$ & $1.83$ & $1.73$  & $1.70$ & $1.73$ &  $1.83$  \\ \hline
    $(G_{+}=0.1) \; q_c$ & $1.78$ & $1.71$  & $1.65$ & $1.71$ &  $1.78$  \\ \hline
    $(G_{\pm}=0.9) \; q_s$ & $1.89$ & $1.87$  & $1.87$ & $1.87$ &  $1.89$  \\ \hline
    $(G_{\pm}=0.5) \; q_s$ & $1.75$ & $1.89$  & $1.88$ & $1.89$ &  $1.75$  \\ \hline
    $(G_{\pm}=0.1) \; q_s$ & $1.85$ & $1.76$  & $1.82$ & $1.94$ &  $1.80$  \\ \hline
    $B^s_{\Gamma}$ & $1.89$ & $1.56$  & $1.46$ & $1.56$ &  $1.89$  \\ \hline
    $ \sqrt{2} B^c_{\Gamma}$ & $1.68$ & $1.27$  & $1.14$ & $1.27$ &  $1.68$  \\ \hline
  \end{tabular}
\end{table}

\begin{figure}[h!]
\centering
		\includegraphics[width=6in]{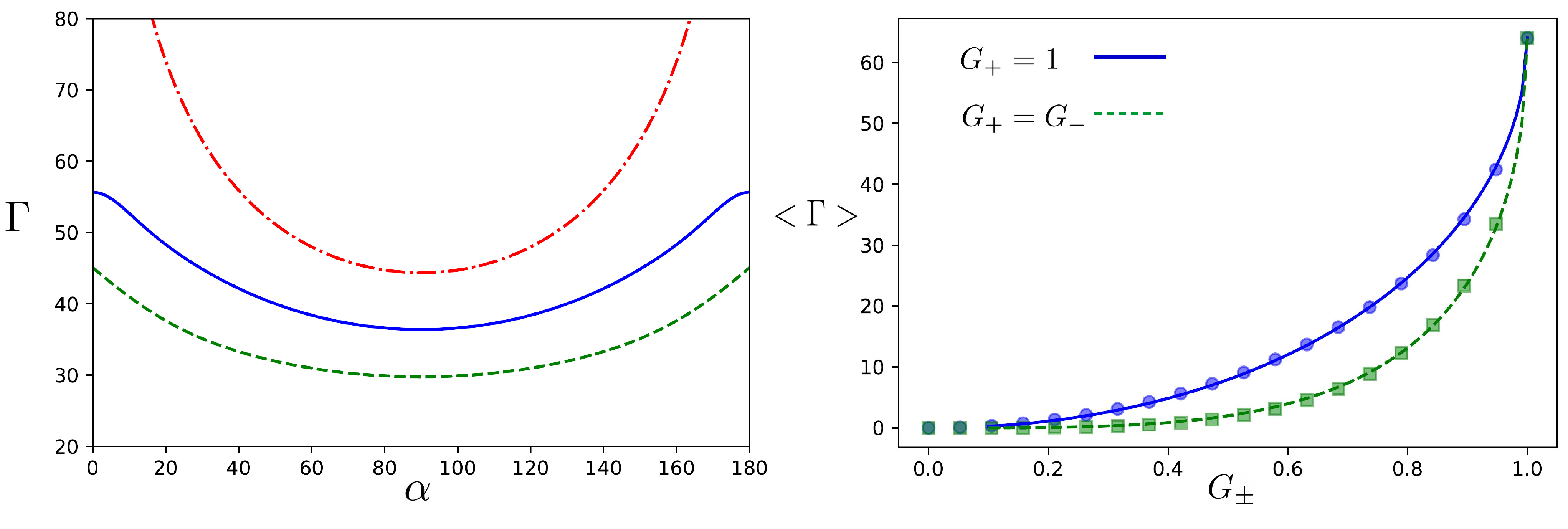}
\caption{Gravitational radiation emission efficiency for cuspless loops. The left panel represents the distribution of $\Gamma$ for loops described by eq.~(\ref{CuspLessLoop}). The dash-dotted line corresponds to the standard case (without current), solid line to the chiral loop with $G_+=1$, $G_-=0.95$ and the dashed line to the symmetric current loop with $G_+=G_-=0.95$. The right panel shows the average values of $\left\langle \Gamma \right\rangle$ fitted using an expression of the form of eq.~(\ref{Gamma_c_s}) for chiral current (solid line) and loops with symmetric current (dashed line) with $B^s_{\Gamma} = 1.73$ and $\sqrt{2} B^c_{\Gamma} = 1.47$.}
\label{fig:ADistrKinkyLoop}
\end{figure}

The efficiency of emission of gravitational radiation $\Gamma$ for cuspless loops with and without current is plotted in the left panel of figure~\ref{fig:ADistrKinkyLoop}, while the dependence of the average value of $\left\langle \Gamma \right\rangle$ on the value of the current is shown on the right panel. These results show that the impact of current on $\Gamma$ for loops with kinks is similar to that on loops with cusps: current causes a decrease in the efficiency of emission of gravitational radiation that is well described by the fitting function in eq.~(\ref{Gamma_c_s}). The fitted values of the parameters of this function are recorded in table~\ref{TableQKink}.

\section{Impact on the stochastic gravitational wave background}\label{sec:SGWB}

In the previous sections, we have shown that currents have a significant impact on the emission of gravitational waves by cosmic string loops, leading not only to a decrease of the power emitted but also changing the emission spectrum of loops with cusps. As a result, the stochastic gravitational wave background (SGWB) generated by a string network --- that results from the superimposition of the transient burst emitted by the many loops created throughout cosmological evolution~\cite{Vilenkin:1981bx} --- may also be significantly affected by the presence of currents. In this section, we will briefly discuss the potential impact of currents on the shape and amplitude of the cosmic strings SGWB.

The SGWB generated by standard (currentless) cosmic strings has been widely studied in the literature~\cite{Vilenkin:1981bx,Binetruy:2012ze,Sanidas:2012ee,Kuroyanagi:2012wm,SousaAvelino,Blanco-Pillado:2013qja,SousaAvelino2,SousaAvelinoGuedes,Blanco-Pillado:2017oxo,LISA}. Its amplitude is generally characterized by the spectral density of gravitational waves,

\begin{equation}
\Omega_{\rm gw}(f)=\frac{1}{\rho_c}\frac{d\rho_{\rm gw}}{d \log{f}}\,,
\end{equation}
in units of critical density of the universe $\rho_c=3H_0^2/(8\pi G)$, where $H_0$ is the value of the Hubble parameter at the present time. This spectrum has, at a given frequency $f$, contributions from all loops that have emitted throughout cosmological history gravitational waves that have a frequency $f$ at the present time. Its amplitude is therefore given by~\cite{Sanidas:2012ee,Blanco-Pillado:2013qja,LISA}:

\begin{equation}
\Omega_{\rm gw}(f)=\frac{16\pi}{3}\left(\frac{G\mu_0}{H_0}\right)^2\frac{1}{f}\int_{t_i}^{t_0}dt'\left(\frac{a(t')}{a(t_0)}\right)^5\sum_{j=1}^{+\infty}j \Gamma_j n(\ell_j(t'),t')\,,
\label{Omega}
\end{equation}
where the integration is carried out from the time in which significant loop production starts $t_i$ (usually identified with the end of the friction-dominated era) until the present time $t_0$, $\ell_j\equiv (2ja(t')/(f a(t_0))$ is the length that a loop should have at a time $t'$ to emit, in the $j$-th harmonic mode, gravitational waves that have a frequency $f$ at $t=t_0$, $\Gamma_j$ is the power emitted in the $j$-th harmonic mode (in $G\mu_0^2$ units) computed in the previous sections, and $a$ is the cosmological scale factor. Here, $n(\ell,t)d\ell$ represents the number density of loops with lengths between $\ell$ and $\ell+d\ell$ that exist at a time $t$.

The impact of decreasing the total power emitted by loops $\Gamma$ on the amplitude of the SGWB is already well understood (see e.g.~\cite{LISA,SousaAvelino}): it leads to an increase of the amplitude of the SGWB, given roughly by $\Omega_{\rm gw}\propto \Gamma^{-1/2}$.  Although this result may seem counter-intuitive at first, it may be simply explained: if we decrease $\Gamma$, loops survive longer, thus emitting their energy in the form of gravitational waves closer to today. The energy density in gravitational waves is therefore less diluted by expansion and an observer today measures a larger $\Omega_{\rm gw}$. However, as eq.~(\ref{Omega}) clearly shows, the emission spectrum is not the only quantity affecting the SGWB: the loop distribution function is another key ingredient in this computations and understanding the impact of currents on $n(\ell,t)$ is pivotal to study their impact on the $\Omega_{\rm gw}$.

Although, in principle, $n(\ell,t)$ can be measured or inferred from numerical simulations --- as was done for Nambu-Goto strings~\cite{Blanco-Pillado:2013qja,Lorenz:2010sm,Blanco-Pillado:2019tbi,SousaAvelinoGuedes} --- no such results currently exist for superconducting strings. Here, therefore, we shall adopt a (semi)-analytical approach to evaluate $n(\ell,t)$, following the framework proposed in~\cite{SousaAvelino}. This approach --- which we only briefly outline here; we refer the reader to~\cite{SousaAvelino,SousaAvelino2,Sousa:2014gka,LISA,SousaAvelinoGuedes} for further details --- resorts to semi-analytical models to describe cosmic string network dynamics to compute the energy that is lost as a result of loop chopping throughout cosmological time. Two variables are, in general, sufficient to describe the evolution of cosmic string networks on sufficiently large scales~\cite{Martins:1996jp}: the characteristic length $L$ --- which is a measure of the energy density of the network $\rho=\mu_0/L^2$ and, for standard strings, of the physical length of strings --- and the root-mean-squared (RMS) velocity $v$. The energy lost into loops is, then, determined by the evolution of these two quantities:

\begin{equation}
    \dot{\rho}_\ell\equiv\left.\frac{d\rho}{dt}\right|_{loops}={\tilde c} \frac{\rho}{L}v\,,
\end{equation}
where ${\tilde c}=0.23$ is a phenomenological parameter quantifying the efficiency of the loop production mechanism that is calibrated with simulations. This framework further assumes that loops are created with a length that is a fixed fraction of the characteristic length of the long string network: $\ell=\alpha L$, with $\alpha<1$\footnote{Although this assumption amounts to assuming that all loops are created with the same length, the results obtained under this assumption may be used to construct $n(\ell,t)$ for any distribution of loop length at the moment of creation, as demonstrated in~\cite{SousaAvelinoGuedes}.}. The loop size parameter $\alpha$ may, in principle, be calibrated with numerical simulations, however, since the length of loops of superconducting strings is unknown, we shall treat it here as free parameter. The number density of loops created then depends merely on the large scale dynamics on the cosmic string network and is given by $\dot{n}_\ell=\dot{\rho}_\ell/(\alpha L)$. Taking into account that the length of loops decreases as a result of gravitational wave emission as $d\ell/dt=-\Gamma G\mu_0$ and that loops are diluted by expansion, $n(\ell,t)$ can then be obtained for any moment in cosmological history and for any $\ell$.

The large-scale dynamics of networks of strings with current was studied in detail in~\cite{MPRS,MPRS2}. Therein, they found that, for fast enough expansion rates, currents are rapidly diluted. As a result, superconducting strings are effectively currentless in the matter era and behave as standard cosmic strings. This means that the contribution of loops created during the matter era to the SGWB remains, unaltered, for superconducting strings. During the radiation era, however, the networks experience a linear scaling regime with non-trivial current characterized by\footnote{Such a scaling regime is only attainable if there is a mechanism to dissipate current --- be it direct leakage or bias~\cite{MPRS}. In the absence of such mechanisms, the networks would evolve towards a non-relativistic regime with growing charge.}:

\begin{equation}
    L_{\rm ph}=\xi_rt\quad\mbox{and}\quad v=v_r\,,
\end{equation}
with

\begin{equation}
\xi_r^2=(1-Y)k(k+{\tilde c})\quad\mbox{and}\quad v_r^2=(1-Y)\frac{k}{k+{\tilde c}}\,.   
\end{equation}
Here, $Y$ is the (averaged) macroscopic charge, related to the microscopic current roughly as $Y\sim \left<F'_\pm\right>^2$, and remains constant during this regime and $k(v)$ is a momentum parameter that takes the form~\cite{Martins:2000cs}:
\begin{equation}
    \label{MomentK}
    k(v) = \frac{2 \sqrt{2}}{\pi} \frac{1-8 v^6}{1+8 v^6}
    \left( 1-v^2 \right) \left( 1 + 2 \sqrt{2} v^3 \right)\,.
\end{equation}
Here, $L_{\rm ph}$ represents the (averaged) physical length of superconducting strings. Note that for strings with current the physical and characteristic lengths no longer coincide: strings have an additional energy in the form of current and charge. As a matter of fact, we have that $L_{\rm ph}=L\sqrt{1+Y}$ and $\rho_0=\mu_0/L_{\rm ph}^2$ may, in this case, be interpreted as the bare string energy density (that does not include the contribution of current). In the remainder of this section we will express all our results in terms of the physical length and bare energy, since these are the relevant quantities to compute the SGWB. Note also that, in the limit $Y\to 0$, we recover the results for Nambu-Goto strings.

\begin{figure}
\begin{center}
\includegraphics[width=6in]{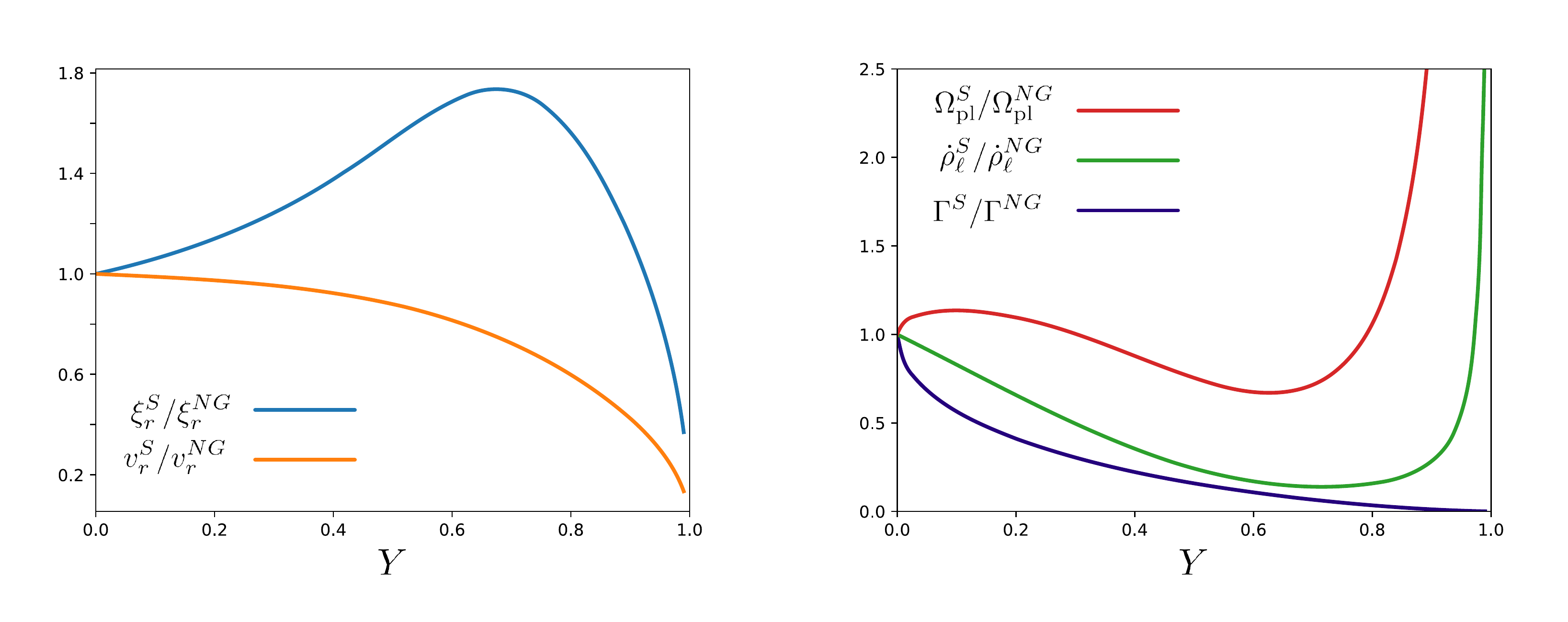}
\caption{\label{figure:ratios} Impact of current on the amplitude of the stochastic gravitational wave background generated by radiation era loops. The left panel shows the impact of the macroscopic current $Y$ on the averaged physical length $\xi_r^S$ and the RMS velocity $v_r^S$ (normalized by the corresponding quantities for Nambu-Goto strings $\xi_r^{NG}$ and $v_r^{NG}$). In the right panel, we plot the ratio between the amplitude of the radiation plateau of the SGWB generated by superconducting and cosmic strings $\Omega_{\rm gw}^S/\Omega_{\rm gw}^{NG}$, as well as the ratio between the energy density lost due to the formation of loops $\dot{\rho}_\ell^S/\dot{\rho}_\ell^{NG}$ and between the total power emitted in gravitational radiation $\Gamma^S/\Gamma^{NG}$.}
\end{center}
\end{figure}

A scaling population of loops decaying in the radiation era gives rise to a flat spectrum and, since networks of strings with current maintain a linear scaling regime during the radiation era, this should also be the case for superconducting strings. However, the amplitude of this plateau should not only be affected by the decrease of gravitational radiation emission efficiency but also by the change in the number of loops produced resulting from the impact of currents on the large-scale network dynamics. Using the analytical results in~\cite{SousaAvelinoGuedes} and assuming that loops are created with a length that is larger than the gravitational radiation scale ($\ell\gg \Gamma G\mu_0$)\footnote{For loops with $\ell\ll \Gamma G\mu_0$, the last term in the expression would not be present.}, we find that:

\begin{equation}
    \frac{\Omega_{\rm pl}^S}{\Omega_{\rm pl}^{NG}}=\frac{\dot{\rho}_\ell^S}{\dot{\rho}_\ell^{NG}}\left(\frac{\Gamma^{NG}\xi_r^S}{\Gamma^S\xi_r^{NG}}\right)^{1/2}\,,
    \label{plateau}
\end{equation}
where $\Omega_{\rm pl}^m$ represents the amplitude of the plateau, $m=S,NG$ labels superconducting and Nambu-Goto strings respectively and we have assumed that superconducting strings and standard strings have the same (bare) tension and create loops characterized by the same $\alpha$ and that $\tilde c$ is independent on the value of the current.
Given the dependency of $k$ on the RMS velocity, the impact of currents on $\xi_r$ and $v_r$ is not trivial; we illustrate this effect in the left panel of figure~\ref{figure:ratios}. Roughly speaking, currents effectively decelerate strings, leading to velocities that are smaller than those of currentless strings. This makes collisions between strings and the production of loops less likely. This trend is further accentuated by the characteristic length of the network, for small enough currents: as $Y$ increases, $\xi_r$ also increases, leading to networks that are more diluted than networks of standard strings and in which collisions are less likely to occur. However, for large enough currents, the networks may become significantly denser then networks of standard strings and production of loops should necessarily be enhanced. In the right panel of Fig.~\ref{figure:ratios} we plot the ratio between the amplitude of the radiation era plateau of the SGWB generated by superconducting and standard cosmic string networks alongside the ratios between the number of loops produced and the values of $\Gamma$. This figure clearly shows that considering both these effects --- the change in the efficiency of emission of gravitational radiation and the impact on the large-scale dynamics --- is essential for an accurate characterization of the amplitude of the gravitational wave background generated by current-carrying strings. In fact, provided that the current is not too large ($Y\lesssim 0.8$), its effect on the amplitude of the contribution of loops created in the radiation era is not very dramatic. Although the gravitational radiation emission efficiency decreases steadily as we increase $Y$, so does the number of loops produced per unit volume and, as a result of the competition between these two effects, there is no significant variation in $\Omega_{\rm pl}$. However, as current continues to grow and $\Gamma\to 0$ and the networks start to become increasingly denser, the amplitude of the radiation era plateau grows very steeply. Note that loops created in the later stages of the radiation era (for $\ell\gg \Gamma G\mu_0$) are expected to survive the radiation-matter transition and partially decay in the matter era. This population of loops gives rise to a peak in the low frequency portion of the spectrum. The amplitude of this contribution should also vary according to eq.~(\ref{plateau}), but it should also shift in frequency according as $f\propto \/\xi_r$.

\begin{figure}
\begin{center}
\includegraphics[width=6in]{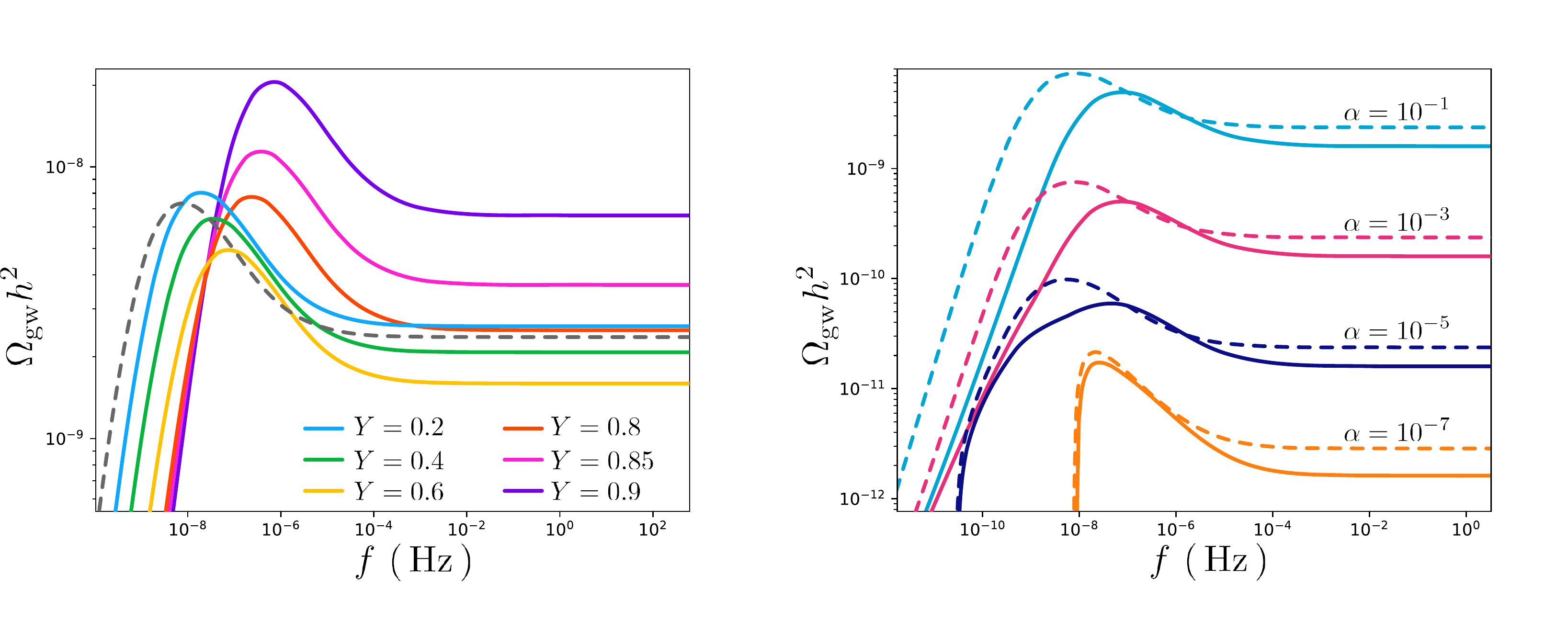}
\caption{Stochastic gravitational wave background generated by networks of superconducting strings. The left panel shows the spectrum generated for loops characterized by $\alpha=10^{-1}$ and different values of macroscopic current $Y$. In the right panel we plot the stochastic gravitational wave background for networks with a macroscopic current of $Y=0.6$ and different values of $\alpha$. We include, in both panels, the spectrum generated by standard strings for comparison (dashed lines). We have considered, both for standard and superconducting strings, loops with kinks, with a power spectrum characterized by the exponent $q=5/3$, and we have included $10^5$ modes of emission. We also took $G\mu_0=10^{-10}$\label{figure:omegas} and $\Gamma^{NG}=50$ and considered chiral currents.}
\end{center}
\end{figure}

In figure~\ref{figure:omegas}, we plot some examples of the SGWB generated by cosmic strings with current alongside that of strings with $Y=0$. For simplicity, we use the analytical approximation in~\cite{SousaAvelinoGuedes} to compute both spectra, since it provides a good fit to the numerically computed spectra. In the left panel, we represent the SGWB for various values of current and $ \alpha = 10^{-1}$. This figure illustrates the discussion of the previous paragraph: although current affects the overall amplitude of the spectrum and causes a shift of the peak towards higher frequencies, these effects are not very significant for $Y\lesssim0.8$. Despite this, there may be situations where the shape of the spectrum is significantly affected. The peak of the spectrum not only has contributions from loops that are created in the radiation era and decay in the matter era, but also from the loops that are created in the matter era. Both these loop populations give rise to peaked contributions to the spectrum, but they have different shapes and amplitudes. So, the shape of the final spectrum depends on the interplay between these two contributions and, since current affects the amplitude of the spectrum created by radiation era loops, this may result in changes to the shape of the peak of the spectrum. Although for large $\alpha$ (and small enough $G \mu_0$), the contribution of loops generated in the radiation era dominates, its amplitude decreases as $\alpha^{1/2}$ while the contribution of matter era is roughly independent of $\alpha$~\cite{SousaAvelinoGuedes}. So, for small enough $\alpha$, changes to the shape of the peak of the spectrum should occur. This may be seen in the right panel of figure~\ref{figure:omegas}, where we plot the SGWB generated by networks of strings with $Y=0.6$ for different values of $\alpha$ alongside the corresponding spectrum for standard cosmic strings. This figure shows that the shape of the spectrum may indeed be significantly different when a current is present for small enough $\alpha$ and the relative height of the peak compared to the plateau portion may also be significantly different too. These differences may be regarded as signatures of current, since they may allow us to distinguish between superconducting strings and Nambu-Goto strings. They may then be regarded as tell-tale signs of superconducting strings.

Note however that these computations of the SGWB generated by cosmic strings with current should only be regarded as preliminary. They have the underlying assumption that the emission of gravitational radiation is the only decay mechanism for superconducting string loops. Loops with current, however, are also expected to emit burst of electromagnetic radiation~\cite{Blanco-PilladoOlumVilenkin} and, as a result, a significant portion of their energy may be lost in the form of electromagnetic waves. Moreover, currents may have an impact on the frequency of the gravitational waves emitted and even prevent the complete collapse of the loops, leading to stable circular string configurations known as vortons~\cite{Davis:1988ij,MartinsShellard1998,CarterPeterGangui}. A detailed computation of the SGWB generated by strings with current would require a complete characterization of the evolution of current on cosmic string loops, including the effect of electromagnetic radiation and possible current leakage mechanisms, which we leave for future work. Including these effects, however, would very likely result in a suppression of the total gravitational wave power emitted throughout the life of cosmic string loops, leading to a decrease of the amplitude of the contribution of radiation era loops to the SGWB. We may, therefore, regard the computations derived here as the largest possible SGWB generated by cosmic strings with current. On the other hand, considering only the contribution of matter era loops may be considered the minimal SGWB generated by superconducting strings. Any more realistic computation would therefore fall within these two limiting scenarios.

\section{Conclusions}\label{sec:conclusions}

We studied how the efficiency of emission of gravitational radiation $\Gamma$ of cosmic string loops changes in the presence of current. In particular, we found that $\Gamma^{m}_{G} = \Gamma_{0} (1-|F^{\prime}_{\pm}|)^{B^{m}_\Gamma}$, where $B^{s}_\Gamma\approx 2$ for loops with cusps and $B^{s}_\Gamma\approx 1.5$ for loops with kinks with symmetrical current, while $B^{s}_\Gamma=\sqrt{2}B^{c}_\Gamma$ in both cases for chiral currents (see figure~\ref{figure:Distrib} and table~\ref{TableQKink} for further details). This shows that the gravitational wave bursts emitted by cosmic string loops with current should be expected to be weaker than those generated by Nambu-Goto strings.

We have also studied the spectrum of emission of gravitational radiation over loop harmonics and found that, although for cuspless loops it is not affected by the inclusion of current and maintains the power-law shape, this is not the case for loops with cusps. As a matter of fact, we have found that, in this case, the spectrum follows to a good approximation an exponential decay with increasing harmonic mode.

We have performed these calculations for a transonic type of current. However, one may see that the result for cuspless loops is valid for any current and the parameters obtained are similar to those of loops with cusps. Hence, we conjecture that the phenomenological relations obtained are justifiable in the general case of current-carrying strings for arbitrary function $\mathcal{F}(\kappa)$. 

These results are essential to study the stochastic gravitational wave background generated by cosmic strings with non-trivial additional degrees of freedom. As a matter of fact, we have performed preliminary studies of the impact of current on the amplitude of the gravitational wave background including the impact of current on $\Gamma$. Our results indicate that, for superconducting strings, the amplitude and shape of the contribution of loops created during the radiation era may differ significantly from that of standard cosmic strings, which may lead to a distinct stochastic gravitational wave background. Note however that, although the inclusion of the decrease of $\Gamma$ caused by current is essential to accurately determine this spectrum, until a proper framework to describe the evolution of current on cosmic string loops is developed --- including the impact of electromagnetic radiation and other leakage mechanisms --- a more precise characterization cannot be performed. We leave the development of such a framework for future work.

\textit{Note:} While this paper was in preparation, related results appeared in~\cite{Auclair:2022ylu}. Therein the authors study the impact of currents on the stochastic gravitational wave background by considering only their impact on the number of loops produced during the radiation era. As this work shows, current leads to a decrease in $\Gamma$ that has a significant impact on the final amplitude of the stochastic gravitational wave background, leading overall to a larger amplitude than that predicted in~\cite{Auclair:2022ylu} (except in $Y=1$ limit, where the emission of gravitational radiation may be completely suppressed).

\appendix

\section{Kaluza-Klein action as a current} \label{Appendix}

The linear action (\ref{Witten}) for superconducting cosmic strings in the Polyakov form may be written as
\begin{equation}
\label{Eff354453}
S_{\text{eff}} = \frac{\mu_0}{2} \int \sqrt{-h} h^{ab} \gamma_{ab} d^2 \sigma - \frac{\mu_0}{2} \int \sqrt{-\gamma} \gamma^{ab} \phi_{,a} \phi_{,b} d^2 \sigma,
\end{equation}
where $h_{ab}$ is a world-sheet dynamical metric, while $\gamma_{ab}=g_{\mu \nu} X^{\mu}_{,a} X^{\nu}_{,b}$ is the independent induced metric. The field $\phi$ ``lives'' on the induced metric in that sense, while if we require that the current evolves on the worldsheet metric $h_{ab}$ and we take only into account the linear contribution from the current, the action takes the form
\begin{equation}
\label{Eff3}
S_{\text{eff}} = \frac{\mu_0}{2} \int \sqrt{-h} h^{ab} \gamma_{ab} d^2 \sigma - \frac{\mu_0}{2} \int \sqrt{-h} h^{ab} \phi_{,a} \phi_{,b} d^2 \sigma.
\end{equation}
By eliminating the auxiliary worldsheet metric $h_{ab}$, one obtains the transonic string action~\cite{CarterSteer}
\begin{equation}
\label{TansActAp}
S_{\text{eff}} = \mu_0 \int \sqrt{1 - \kappa} \sqrt{-\gamma}  d^2 \sigma.
\end{equation}

\section{Useful analytic expressions for the integrals} 
\subsection{Airy functions} \label{Appendix2}

The $I^\mu_\pm$ integrals in section~\ref{Cusp-like point} have the following form
\begin{equation}
\label{IntegralApp}
\mathcal{I} =  \int_{-\infty}^{\infty} (a+b \sigma) \text{e}^{i \omega \left( \alpha \sigma + \beta \sigma^2 + \gamma \sigma^3 \right)} d \sigma,
\end{equation}
(where all variables $\in \Re$) and they may be represented by the following expression
\begin{equation}
\label{IntegralApp2}
\mathcal{I} = 2 B \pi \frac{\gamma \omega}{|\gamma \omega|} \text{e}^{i \omega d} \left( (a+b A) \text{Ai}(\zeta) - i b B \text{Ai}^{\prime}(\zeta) \right),
\end{equation}
where $\text{Ai}(..)$ and $\text{Ai}^{\prime}(..)$ are the Airy function and its derivative respectively,
while  
\begin{equation}
\begin{gathered}
d = \beta \frac{2 \beta^2 - 9 \alpha \gamma}{27 \gamma^2}, \quad A = - \frac{\beta}{3 \gamma}, \\
B = \frac{\gamma \omega}{|\gamma \omega| (3 |\gamma \omega|)^{1/3}}, \quad \zeta = - \frac{\gamma \omega}{|\gamma \omega|} \frac{(\beta^2 - 3 \alpha \gamma) \omega}{3 \gamma (3|\gamma \omega|)^{1/3}}.
\end{gathered}    
\end{equation}
The radiation emitted by cusps has been usually expressed in terms of modified Bessel functions of order $1/3$ (see refs.~\cite{SpergelPiranGoodman, BlancoPilladoOlum,  BabichevDokuchaev2}). Unfortunately, the expression with Bessel functions fails to represent integral (\ref{IntegralApp}) correctly when the argument is negative. Airy functions, however, do not have this drawback\footnote{``It is easier, however, to use the Airy integral directly, and Bessel functions of order $1/3$ seem to have no application except to provide an inconvenient way of expressing this function'' \cite{Jeffreys}.}.

The generalized expression of the integral has the form:
\begin{equation}
\label{IntegrAir}
\int_{-\infty}^{\infty} \sigma^n \text{e}^{i \omega \left( \alpha \sigma + \beta \sigma^2 + \gamma \sigma^3 \right)} d \sigma = \text{e}^{i \omega d} B A^{n} \sum_{k=0}^n \binom{n}{k} \left( \frac{B}{A} \right)^k \text{Ai}^{(k)}(\zeta),
\end{equation}
where $n \in \textbf{Z}^+$.

\subsection{Bessel functions}
\label{Loops integrals}

The $I^\mu_\pm$ integrals in section~\ref{sec:burden} satisfy the following relations
\begin{equation}
\begin{gathered}
\label{InegrLoop1}
\int_0^{2 \pi} \text{e}^{-i N \left( x - a \cos M x - b \sin M x \right)} \begin{cases} \sin M x \\ \cos M x \\ 1 \end{cases} d x = \\ 
= \begin{cases}
  \frac{\pi}{i} \left( \text{e}^{i \delta p_1} J_{p_1}(N d) - \text{e}^{i \delta p_2} J_{p_2}(N d) \right), \quad \text{if: } \; N \mid M \\
    \pi \left( \text{e}^{i \delta p_1} J_{p_1}(N d) + \text{e}^{i \delta p_2} J_{p_2}(N d) \right), \quad \text{if: } \; N \mid M, \\
  2 \pi \text{e}^{i \delta \frac{N}{M}} J_{\frac{N}{M}}(N d), \quad \text{if: } \; N \mid M      
\end{cases}
\end{gathered}
\end{equation}
where $p_1 = \frac{N}{M}-1$, $p_2 = \frac{N}{M}+1$, $\delta = \arctan \frac{a}{b}$ and $d = \text{sign}(b) \sqrt{a^2+b^2}$, and integral vanishes for other choices of $N$ and $M$. See appendix of ref.~\cite{BabichevDokuchaev2002} for more details.

\acknowledgments
We thank Pierre Auclair for useful discussions. L. S. is supported by FCT - Funda\c{c}\~{a}o para a Ci\^{e}ncia e a Tecnologia through contract No. DL 57/2016/CP1364/CT0001. Funding for this work has also been provided by FCT through national funds (PTDC/FIS-PAR/31938/2017) and by FEDER—Fundo Europeu de Desenvolvimento Regional through COMPETE2020 - Programme for Competitiveness and Internationalisation (POCI-01-0145-FEDER-031938), and through the research grants UIDB/04434/2020 and UIDP/04434/2020.



\begin{thebibliography}{99}


\bibitem{Kibble}
T. W. B. Kibble, \emph{Topology of cosmic domains and strings}, \emph{J.Phys. A} {\bf 9} (8)
(1976) 1387.

\bibitem{SarangiTye}
S. Sarangi, S.-H. H. Tye, \emph{Cosmic string production towards the end of brane inflation}, \emph{Phys.Lett.B} {\bf 536} (3-4) (2002) 185 
arXiv:hep-th/0204074.

\bibitem{FirouzjahiTye}
H. Firouzjahi, S.-H. H. Tye, \emph{Brane inflation and cosmic string tension in superstring theory}, \emph{JCAP} {\bf 0503} (2005) 009 
arXiv:hep-th/0501099v3.

\bibitem{LazaridesPeddieVamvasakis}
G. Lazarides, I. N. R. Peddie, A. Vamvasakis, \emph{Semi-shifted hybrid inflation with B-L cosmic strings}, \emph{Phys.Rev. D} {\bf (78)} (2008) 043518 
arXiv:0804.3661v2.

\bibitem{JonesStoicaTye}
N. T. Jones, H. Stoica, S.-H. H. Tye, \emph{The production, spectrum and evolution of cosmic strings in brane inflation}, \emph{Phys.Lett. B} {\bf (563)} (2003) 6–14
arXiv:hep-th/0303269v1.

\bibitem{ChernoffTye}
D. F. Chernoff, S.-H. H. Tye, \emph{Inflation, string theory and cosmic strings},\emph{Int.J.Mod.Phys.D} {\bf 24} (3) (2015) 1530010 
arXiv:1412.0579.

\bibitem{DavisDavisTrodden}
S. C. Davis, A. C. Davis, M. Trodden, \emph{$N=1$ supersymmetric cosmic strings}, \emph{Phys.Lett. B} {\bf 405} (1997) 257–264
arXiv:hep-ph/9702360.


\bibitem{JeannerotRocherSakellariadou}
R. Jeannerot, J. Rocher, M. Sakellariadou, \emph{How generic is cosmic string formation in supersymmetric grand unified theories}, \emph{Phys.Rev.D} {\bf 68} (2003) 103514 
arXiv:hep-ph/0308134

\bibitem{CuiMartinMorrisseyWells}
Y. Cui, S. P. Martin, D. E. Morrissey, J. D. Wells, \emph{Cosmic strings from supersymmetric flat directions}, \emph{Phys.Rev.D} {\bf 77} (2008) 043528
arXiv:0709.0950v2.

\bibitem{Allys}
E. Allys, \emph{Bosonic structure of realistic SO(10) supersymmetric cosmic strings}, \emph{Phys.Rev.D} {\bf 93} (10) (2016) 105021 
arXiv:1512.02029

\bibitem{Allys2}
E. Allys, \emph{Bosonic condensates in realistic supersymmetric GUT cosmic strings}, \emph{Phys.Rev.D} {\bf 93} (10) (2016) 105021
arXiv:1512.02029

\bibitem{DAVIS1986225}
R. Davis, \emph{Cosmic axions from cosmic strings}, \emph{Phys.Lett.B} {\bf 180} (3) (1986) 225–230.

\bibitem{DABHOLKAR1990815}
A. Dabholkar, J. M. Quashnock, \emph{Pinning down the axion}, \emph{Nucl.Phys.B} {\bf 333} (3) (1990) 815–832.

\bibitem{GorghettoHardyVilladoro}
M. Gorghetto, E. Hardy, G. Villadoro, \emph{Axions from strings: the attractive solution}, \emph{JHEP} {\bf 2018} (2018) 
arXiv:1806.04677.

\bibitem{KawasakiKen'ichiSekiguchi}
M. Kawasaki, K. Saikawa, T. Sekiguchi, \emph{Axion dark matter from topological defects}, \emph{Phys.Rev.D} {\bf 91} (2015) 065014
arXiv:1412.0789v3.

\bibitem{PhysRevLett.124.041804}
J. A. Dror, T. Hiramatsu, K. Kohri, H. Murayama, G. White, \emph{Testing the seesaw mechanism and leptogenesis with gravitational waves}, \emph{Phys.Rev.Lett.} {\bf 124} (2020) 041804
arXiv:1908.03227

\bibitem{Samanta2021}
R. Samanta, S. Datta, \emph{Gravitational wave complementarity and impact of nanograv data on gravitational leptogenesis}, \emph{JHEP} {\bf 2021} (5) (2021) 211
arXiv:2009.13452.

\bibitem{LazanauShellard}
A. Lazanu, E. P. S. Shellard, \emph{Constraints on the nambu-goto cosmic string contribution to the CMB power spectrum in light of new temperature and polarisation data}, \emph{JCAP} {\bf 2015} (02) (2015) 024 arXiv:1410.5046v3.

\bibitem{LazanuShellardLandriau}
A. Lazanu, E. P. S. Shellard, M. Landriau, \emph{CMB power spectrum of Nambu-Goto cosmic strings}, \emph{Phys.Rev.D} {\bf 91} (2015) 083519
arXiv:1410.4860v3

\bibitem{LizarragaUrrestillaDaverioHindmarshKunz}
J. Lizarraga, J. Urrestilla, D. Daverio, M. Hindmarsh, M. Kunz, \emph{New CMB constraints for Abelian Higgs cosmic strings}, \emph{JCAP} {\bf 1610} (10) (2016) 042
arXiv:1609.03386v3.

\bibitem{CharnockAvgoustidisCopelandMoss}
T. Charnock, A. Avgoustidis, E. Copeland, A. Moss, \emph{CMB constraints on cosmic strings and superstrings}, \emph{Phys.Rev.D} {\bf 93} (12) (2016) 123503
arXiv:1603.01275.

\bibitem{RybakAvgoustidisMartins}
I. Yu. Rybak, A. Avgoustidis, C. J. A. P. Martins, \emph{Semianalytic calculation of cosmic microwave background anisotropies from wiggly and superconducting cosmic strings}, \emph{Phys.Rev.D} {\bf 96} (10) (2017) 103535 [\emph{Erratum ibid} {\bf 100} (2019) 049901]
arXiv:1709.01839

\bibitem{RybakSousa}
I. Yu. Rybak, L. Sousa, \emph{CMB anisotropies generated by cosmic string loops}, \emph{Phys.Rev.D} {\bf 104} (2021) 023507 
arXiv:2104.08375v2

\bibitem{Sazhin1}
M. V. Sazhin, O. S. Khovanskaya, M. Capaccioli, G. Longo, M. Paolillo, G. Covone, N. A. Grogin, E. J. Schreier, \emph{Gravitational lensing by cosmic strings: what we learn from the CSL-1 case}, \emph{Mon. Not. R. Astron. Soc.} {\bf 376} (4) (2007) 1731–1739 
arXiv:0611744v2.

\bibitem{Sazhin2}
O. S. Sazhina, D. Scognamiglio, M. V. Sazhin, M. Capaccioli, \emph{Optical analysis of a CMB cosmic string candidate}, \emph{Mon. Not. R. Astron. Soc.} {\bf 485} (2) (2019) 1876–1885 
arXiv:1902.08156v1.

\bibitem{SousaAvelino}
L. Sousa, P. P. Avelino, \emph{Stochastic gravitational wave background generated by cosmic string networks: Velocity-dependent one-scale model versus scale-invariant evolution}, \emph{Phys.Rev.D} {\bf 88} (2013) 023516 
arXiv:1403.2621.

\bibitem{Blanco-PilladoOlum}
J. J. Blanco-Pillado, K. D. Olum, \emph{Stochastic gravitational wave background from smoothed cosmic string loops}, \emph{Phys.Rev.D} {\bf 96} (2017) 104046
arXiv:1709.02693v2.

\bibitem{RingevalSuyama}
C. Ringeval, T. Suyama, \emph{Stochastic gravitational waves from cosmic string loops in scaling}, \emph{JCAP} {\bf 2017} (12) (2017) 027–027 
arXiv:1709.03845.

\bibitem{SousaAvelino2}
L. Sousa, P. P. Avelino, \emph{Probing cosmic superstrings with gravitational waves}, \emph{Phys.Rev.D} {\bf 94} (2016) 063529
arXiv:1606.05585

\bibitem{LISA}
P. Auclair, J. J. Blanco-Pillado, D. G. Figueroa4, A. C. Jenkins, M. Lewicki, M. Sakellariadou, S. Sanidas, L. Sousa, D. A. Steer, J. M. Wachter and S. Kuroyanagi, \emph{Probing the gravitational wave background from cosmic strings with LISA}, \emph{JCAP} {\bf 04} (2020) 034
arXiv:1909.00819

\bibitem{Babul:1987me}
A. Babul, T. Piran, D. N. Spergel, \emph{Bosonic superconducting cosmic strings. 1. Classical field theory solutions}, \emph{Phys.Lett.B} {\bf 202} (1988) 307–314.

\bibitem{Everett}
A. E. Everett, \emph{New mechansim for superconductivity in cosmic strings}, \emph{Phys.Rev.Lett.} {\bf 61} (1988) 1807–1810.

\bibitem{DavisPerkins}
A.-C. Davis, W. B. Perkins, \emph{Generic current-carrying strings}, \emph{Phys.Lett.B} {\bf 390} (1) (1997) 107 – 114
arXiv:hep-ph/9610292

\bibitem{Peter:1993tm}
P. Peter, \emph{Spontaneous current generation in cosmic strings}, \emph{Phys.Rev.D} {\bf 49} (1994) 5052–5062 arXiv:hep-ph/9312280.

\bibitem{GaraudVolkov}
J. Garaud, M. S. Volkov, \emph{Superconducting non-Abelian vortices in Weinberg–Salam theory – electroweak thunderbolts}, \emph{Nucl.Phys.B} {\bf 826} (1) (2010) 174 – 216
arXiv:0906.2996.

\bibitem{DavisPeter}
A.-N. Davis, P. Peter, \emph{Cosmic strings are current-carrying}, \emph{Phys.Lett.B} {\bf 358} (3) (1995) 197 – 202
arXiv:hep-ph/9506433v1.

\bibitem{Lilley:2010av}
M. Lilley, F. Di Marco, J. Martin, P. Peter, \emph{Nonabelian Bosonic Currents in Cosmic Strings}, \emph{Phys.Rev.D} {\bf 82} (2010) 023510
arXiv:1003.4601.

\bibitem{Fukuda:2020kym}
H. Fukuda, A. V. Manohar, H. Murayama, O. Telem, \emph{Axion strings are superconducting}, \emph{JHEP} {\bf 2021} (6) (2021) 52
arXiv:2010.02763

\bibitem{AbeHamadaYoshioka}
Y. Abe, Y. Hamada, K. Yoshioka, \emph{Electroweak axion string and superconductivity}, \emph{JHEP} {\bf 2021} (6) (2021) 172
arXiv:2010.02834.

\bibitem{BlancoPilladoOlum}
Blanco-Pillado, K. D. Olum, \emph{Electromagnetic radiation from superconducting string cusps}, \emph{Nucl.Phys.B} {\bf 599} (1) (2001) 435–445
arXiv:astro-ph/0008297.

\bibitem{MiyamotoNakayama}
K. Miyamoto, K. Nakayama, \emph{Cosmological and astrophysical constraints on superconducting cosmic strings}, \emph{JCAP} {\bf 2013} (07) (2013) 012–012 
arXiv:1212.6687.

\bibitem{Imtiaz2020}
B. Imtiaz, R. Shi, Y.-F. Cai, \emph{Updated constraints on superconducting cosmic strings from the astronomy of fast radio bursts}, \emph{Eur.Phys.J.C} {\bf 80} (6) (2020) 500
arXiv:2001.11149

\bibitem{BlinnikovKhlopov}
S. Blinnikov, M. Khlopov, \emph{Possible astronomical effects of mirror particles}, \emph{Soviet Astronomy} {\bf 27} (1983) 371.

\bibitem{SazhinKhlopov}
M. Sazhin, M. Khlopov, \emph{Cosmic strings and gravitational lens effects}, \emph{Soviet Astronomy} {\bf 33} (1989) 98.

\bibitem{BerezinskyVilenkin}
V.Berezinsky, A.Vilenkin, \emph{Ultra high energy neutrinos from hidden-sector topological defects}, \emph{Phys.Rev.D} {\bf 62} (2000) 083512
arXiv:hep-ph/9908257

\bibitem{HydeLongVachaspati}
J. M. Hyde, A. J. Long, T. Vachaspati, \emph{Dark strings and their couplings to the standard model}, \emph{Phys.Rev.D} {\bf 89} (2014) 065031 
arXiv:1312.4573

\bibitem{LongHydeVachaspati}
A. J. Long, J. M. Hyde, T. Vachaspati, \emph{Cosmic strings in hidden sectors: 1. radiation of standard model particles}, \emph{JCAP} {\bf 2014} (09) (2014) 030–030
arXiv:1405.7679

\bibitem{LongVachaspati}
A. J. Long, T. Vachaspati, \emph{Cosmic strings in hidden sectors: 2. cosmological and astrophysical signatures}, \emph{JCAP} {\bf 2014} (12) (2014) 040–040
arXiv:1409.6979

\bibitem{MPRS}
C. J. A. P. Martins, P. Peter, I. Y. Rybak, E. P. S. Shellard, \emph{Generalized velocity-dependent one-scale model for current-carrying strings}, \emph{Phys.Rev.D} {\bf 103} (2021) 043538
arXiv:2011.09700v1

\bibitem{MPRS2}
C. J. A. P. Martins, P. Peter, I. Y. Rybak, E. P. S. Shellard, \emph{Charge-velocity-dependent one-scale linear model}, \emph{Phys.Rev.D} {\bf 104} (2021) 103506
arXiv:2108.03147v2.

\bibitem{Witten84}
E. Witten, \emph{Superconducting Strings}, \emph{Nucl.Phys.B} {\bf 249} (1985) 557–592.

\bibitem{Carter:1989dp}
B. Carter, \emph{Duality Relation Between Charged Elastic Strings and Superconducting Cosmic Strings}, \emph{Phys.Lett.B} {\bf 224} (1989) 61–66.

\bibitem{CARTER1989}
B. Carter, \emph{Stability and characteristic propagation speeds in superconducting cosmic and other string models}, \emph{Phys.Lett.B} {\bf 228} (4) (1989) 466 – 470.

\bibitem{Carter90}
B. Carter, \emph{Integrable equation of state for noisy cosmic string}, \emph{Phys.Rev.D} {\bf 41} (1990) 3869–3872.

\bibitem{Carter95}
B. Carter, \emph{Transonic elastic model for wiggly Goto-Nambu string}, \emph{Phys.Rev.Lett.} {\bf 74} (1995) 3098–3101
arXiv:hep-th/9411231v1.

\bibitem{CarterPeter}
B. Carter, P. Peter, \emph{Supersonic string models for witten vortices}, \emph{Phys.Rev.D} {\bf 52} (1995) R1744–R1748
arXiv:hep-ph/9411425v1.

\bibitem{Carter2000}
B. Carter, \emph{Dilatonic formulation for conducting cosmic string models}, \emph{Ann.Phys.} {\bf 9} (3-5) (2000) 247–257 
arXiv:hep-th/0002162v1.

\bibitem{Vilenkin90}
A. Vilenkin, \emph{Effect of Small Scale Structure on the Dynamics of Cosmic Strings}, \emph{Phys.Rev.D} {\bf 41} (1990) 3038.

\bibitem{NielsenOlesen}
N. Nielsen, P. Olesen, \emph{Dynamical properties of superconducting cosmic strings}, \emph{Nucl.Phys.B} {\bf 291} (1987) 829 – 846.

\bibitem{Nielsen}
N. Nielsen, \emph{Dimensional reduction and classical strings}, \emph{Nucl.Phys.B} {\bf 167} (1) (1980) 249 – 260.

\bibitem{Carter2}
B. Carter, \emph{Brane dynamics for treatment of cosmic strings and vortons}, \emph{2nd Mexican School on Gravitation and Mathematical Physics} (1997) 
arXiv:hep-th/9705172.

\bibitem{CarterSteer}
B. Carter, D. A. Steer, \emph{Symplectic structure for elastic and chiral conducting cosmic string models}, \emph{Phys.Rev.D} {\bf 69} (2004) 125002
arXiv:hep-th/0307161.

\bibitem{RybakAvgoustidisMartins2}
I. Y. Rybak, A. Avgoustidis, C. J. A. P. Martins, \emph{Collisions of cosmic strings with chiral currents}, \emph{Phys.Rev.D} {\bf 98} (2018) 063519 
arXiv:1809.04033v1

\bibitem{Rybak}
I. Y. Rybak, \emph{Revisiting y junctions for strings with currents: Transonic elastic case}, \emph{Phys.Rev.D} {\bf 102} (2020) 083516
arXiv:arXiv:2001.07262v3

\bibitem{AllenShellard1992}
B. Allen, E. P. S. Shellard, \emph{Gravitational radiation from cosmic strings}, \emph{Phys.Rev.D} {\bf 45} (1992) 1898–1912.

\bibitem{Durrer}
R. Durrer, \emph{Gravitational angular momentum radiation of cosmic strings}, \emph{Nucl.Phys.B} {\bf 328} (1) (1989) 238–271.

\bibitem{GarfinkleVachaspati}
D. Garfinkle, T. Vachaspati, \emph{Fields due to kinky, cuspless, cosmic loops}, \emph{Phys.Rev.D} {\bf 37} (1988) 257–262.

\bibitem{SpergelPiranGoodman}
D. N. Spergel, T. Piran, J. Goodman, \emph{Dynamics of superconducting cosmic strings}, \emph{Nucl.Phys.B} {\bf 291} (1987) 847–875.

\bibitem{DamourVilenkin2}
T. Damour, A. Vilenkin, \emph{Gravitational wave bursts from cosmic strings}, \emph{Phys.Rev.Lett.} {\bf 85} (2000) 3761–3764 
arXiv:gr-qc/0004075.

\bibitem{DamourVilenkin}
T. Damour, A. Vilenkin, \emph{Gravitational wave bursts from cusps and kinks on cosmic strings}, \emph{Phys.Rev.D} {\bf 64} (2001) 064008 
arXiv:gr-qc/0104026.

\bibitem{BabichevDokuchaev2}
E. Babichev, V. Dokuchaev, \emph{Gravitational radiation from chiral string cusps}, \emph{Phys.Rev.D} {\bf 67} (2003) 125016
arXiv:astro-ph/0303659.

\bibitem{BURDEN1985}
C. Burden, \emph{Gravitational radiation from a particular class of cosmic strings}, \emph{Phys.Lett.B} {\bf 164} (4) (1985) 277–281.

\bibitem{BabichevDokuchaev2002}
E. Babichev, V. Dokuchaev, \emph{Oscillation damping of chiral string loops}, \emph{Phys.Rev.D} {\bf 66} (2002) 025007 
arXiv:hep-ph/0204304.

\bibitem{VachaspatiVilenkin}
T. Vachaspati, A. Vilenkin, \emph{Gravitational radiation from cosmic strings}, \emph{Phys.Rev.D} {\bf 31} (1985) 3052–3058.

\bibitem{Vilenkin:2000jqa}
A. Vilenkin, E. P. S. Shellard, \emph{Cosmic Strings and Other Topological Defects}, Cambridge University Press, (2000), pg.517.

\bibitem{GarfinkleVachaspati2}
D. Garfinkle, T. Vachaspati, \emph{Radiation from kinky, cuspless cosmic loops}, \emph{Phys.Rev.D} {\bf 36} (1987) 2229–2241.

\bibitem{CopelandHindmarshTurok}
E. Copeland, D. Haws, M. Hindmarsh, N. Turok, \emph{Dynamics of and radiation from superconducting strings and springs}, \emph{Nucl.Phys.B} {\bf 306} (4) (1988) 908–930.

\bibitem{Vilenkin:1981bx}
A. Vilenkin, \emph{Gravitational radiation from cosmic strings}, \emph{Phys.Lett.B} {\bf 107} (1981) 47–50.

\bibitem{Binetruy:2012ze}
P. Binetruy, A. Bohe, C. Caprini, J.-F. Dufaux, \emph{Cosmological Backgrounds of Gravitational Waves and eLISA/NGO: Phase Transitions, Cosmic Strings and Other Sources}, \emph{JCAP} {\bf 06} (2012) 027 
arXiv:1201.0983.

\bibitem{Sanidas:2012ee}
S. A. Sanidas, R. A. Battye, B. W. Stappers, \emph{Constraints on cosmic string tension imposed by the limit on the stochastic gravitational wave back ground from the European Pulsar Timing Array}, \emph{Phys.Rev.D} {\bf 85} (2012) 122003 
arXiv:1201.2419.

\bibitem{Kuroyanagi:2012wm}
S. Kuroyanagi, K. Miyamoto, T. Sekiguchi, K. Takahashi, J. Silk, \emph{Forecast constraints on cosmic string parameters from gravitational wave direct detection experiments}, \emph{Phys.Rev.D} {\bf 86} (2012) 023503 
arXiv:1202.3032.

\bibitem{Blanco-Pillado:2013qja}
J. J. Blanco-Pillado, K. D. Olum, B. Shlaer, \emph{The number of cosmic string loops}, \emph{Phys.Rev.D} {\bf 89} (2) (2014) 023512 
arXiv:1309.6637.

\bibitem{SousaAvelinoGuedes}
L. Sousa, P. P. Avelino, G. S. F. Guedes, \emph{Full analytical approximation to the stochastic gravitational wave background generated by cosmic string networks}, \emph{Phys.Rev.D} {\bf 101} (2020) 103508 
arXiv:2002.01079v1

\bibitem{Blanco-Pillado:2017oxo}
J. J. Blanco-Pillado, K. D. Olum, \emph{Stochastic gravitational wave background from smoothed cosmic string loops}, \emph{Phys.Rev.D} {\bf 96} (10) (2017) 104046
arXiv:1709.02693

\bibitem{Lorenz:2010sm}
L. Lorenz, C. Ringeval, M. Sakellariadou, \emph{Cosmic string loop distribution on all length scales and at any redshift}, \emph{JCAP} {\bf 10} (2010) 003 
arXiv:1006.0931.

\bibitem{Blanco-Pillado:2019tbi}
J. J. Blanco-Pillado, K. D. Olum, \emph{Direct determination of cosmic string loop density from simulations}, \emph{Phys.Rev.D} {\bf 101} (10) (2020) 103018 
arXiv:1912.10017.

\bibitem{Sousa:2014gka}
L. Sousa, P. P. Avelino, \emph{Stochastic gravitational wave background generated by cosmic string networks: The small-loop regime}, \emph{Phys.Rev.D} {\bf 89} (8) (2014) 083503 
arXiv:1403.2621.

\bibitem{Martins:1996jp}
C. J. A. P. Martins, E. P. S. Shellard, \emph{Quantitative string evolution}, \emph{Phys.Rev.D} {\bf 54} (1996) 2535–2556. 
arXiv:hep-ph/9602271

\bibitem{Martins:2000cs}
C. J. A. P. Martins, E. P. S. Shellard, \emph{Extending the velocity dependent one scale string evolution model}, \emph{Phys.Rev.D} {\bf 65} (2002) 043514 
arXiv:hep-ph/0003298.

\bibitem{Blanco-PilladoOlumVilenkin}
J. J. Blanco-Pillado, K. D. Olum, A. Vilenkin, \emph{Dynamics of superconducting strings with chiral currents}, \emph{Phys.Rev.D} {\bf 63} (2001) 103513.
arXiv:astro-ph/0004410v3

\bibitem{Davis:1988ij}
R. L. Davis, E. P. S. Shellard, \emph{Cosmic vortons}, \emph{Nucl.Phys.B} {\bf 323} (1989) 209–224.

\bibitem{MartinsShellard1998}
C. Martins, E. Shellard, \emph{Vorton formation}, \emph{Phys.Rev.D} {\bf 57} (1998) 7155–7176 
arXiv:hep-ph/9804378

\bibitem{CarterPeterGangui}
B. Carter, P. Peter and A. Gangui, \emph{Avoidance of collapse by circular current-carrying cosmic string loops}, \emph{Phys.Rev.D} {\bf 55} (8) (1997) 4647-4662
arXiv:hep-ph/9609401v1.

\bibitem{Auclair:2022ylu}
P. Auclair, S. Blasi, V. Brdar, K. Schmitz, \emph{Gravitational Waves from Current-Carrying Cosmic Strings} (2022) 
arXiv:2207.03510

\bibitem{Jeffreys}
H. J. F.R.S., \emph{XLIV. Asymptotic solutions of linear differential equations}, \emph{The London, Edinburgh, and Dublin Philosophical Magazine and Journal of Science} {\bf 33} (221) (1942) 451–456.







\end{thebibliography}
\end{document}